\documentclass[
 reprint,
 amsmath,amssymb,
 aps,
 superscriptaddress
]{revtex4-2}

\usepackage{graphicx}
\usepackage{dcolumn}
\usepackage{bm}
\usepackage{braket}
\usepackage{xcolor}
\usepackage{caption}
\usepackage{subcaption}
\usepackage[normalem]{ulem}

\newcommand{\argmin}[1]{\underset{#1}{\mathrm{argmin}}}
\newcommand{\argmax}[1]{\underset{#1}{\mathrm{argmax}}}
\newcommand{\tr}{\mathrm{Tr}}
\renewcommand{\vec}{\bm}

\begin{document}

\preprint{APS/123-QED}

\title{Variational Quantum Time Evolution without the Quantum Geometric Tensor}

\author{Julien Gacon}
\affiliation{%
    IBM Quantum, IBM Research Europe – Zurich, CH-8803 Rüschlikon, Switzerland
}
\affiliation{%
    Institute of Physics, École Polytechnique Fédérale de Lausanne (EPFL), CH-1015 Lausanne, Switzerland
}
\author{Jannes Nys}
\affiliation{%
    Institute of Physics, École Polytechnique Fédérale de Lausanne (EPFL), CH-1015 Lausanne, Switzerland
}
\author{Riccardo Rossi}
\affiliation{%
    Sorbonne Universit\'e, CNRS, Laboratoire de Physique Th\'eorique de la Mati\`ere Condens\'ee, LPTMC, F-75005 Paris, France
}
\author{Stefan Woerner}
\affiliation{%
    IBM Quantum, IBM Research Europe – Zurich, CH-8803 Rüschlikon, Switzerland
}
\author{Giuseppe Carleo}
\affiliation{%
    Institute of Physics, École Polytechnique Fédérale de Lausanne (EPFL), CH-1015 Lausanne, Switzerland
}

\date{\today}

\begin{abstract}
The real- and imaginary-time evolution of quantum states are powerful tools in physics, chemistry, and beyond, to investigate quantum dynamics, prepare ground states or calculate thermodynamic observables.
On near-term devices, variational quantum time evolution is a promising candidate for these tasks, as the required circuit model can be tailored to trade off available device capabilities and approximation accuracy.
However, even if the circuits can be reliably executed, variational quantum time evolution algorithms quickly become infeasible for relevant system sizes due to the calculation of the Quantum Geometric Tensor (QGT).
In this work, we propose a solution to this scaling problem by leveraging a dual formulation that circumvents the explicit evaluation of the QGT.
We demonstrate our algorithm for the time evolution of the Heisenberg Hamiltonian and show that it accurately reproduces the system dynamics at a fraction of the cost of standard variational quantum time evolution algorithms. As an application of quantum imaginary-time evolution, we calculate a thermodynamic observable, the energy per site, of the Heisenberg model.
\end{abstract}

\maketitle

\section{Introduction}
\label{sec:introduction}

Quantum time evolution is a central task in physics. Real-time evolution provides detailed insight into properties of quantum mechanical systems, such as phase transitions~\cite{zhang_transitions_2017, dborin_gs_and_transitions_2022, ebadi_256atoms_2021} or thermalization~\cite{altman_thermalization_2018, dejong_thermalization_2022}.
Imaginary-time evolution is an important tool that enables the preparation of ground states or thermal states \cite{mcardle_variational_2019, jones_vqd_2019, motta_determining_2020}. These can, in turn, be used for the calculation of thermodynamic observables \cite{motta_determining_2020, getelina_qmetts_2023}. In particular, combining real- and imaginary-time evolution would allow the direct calculation of dynamical correlation functions at thermal equilibrium.

The range of applications of imaginary-time evolution extends beyond the field of physics.
Ground-state preparation with imaginary-time evolution for gapped, non-degenerate Hamiltonians is guaranteed to converge in the generic case of non-zero overlap between the ground state and the initial trial state. This makes it a promising candidate in settings where a good initial state can be constructed, e.g.\ in chemistry applications~\cite{barkoutsos_electronic_2018} or in classical optimization problems~\cite{zoufal_blackbox_2023}. 
In quantum machine learning, the preparation of Gibbs states with imaginary-time evolution is a subroutine for quantum Boltzmann machines, which can, for example, be used in distribution learning or classification~\cite{zoufal_variational_2021}.

\begin{figure}[t]
    \centering
    \includegraphics[width=0.9\linewidth]{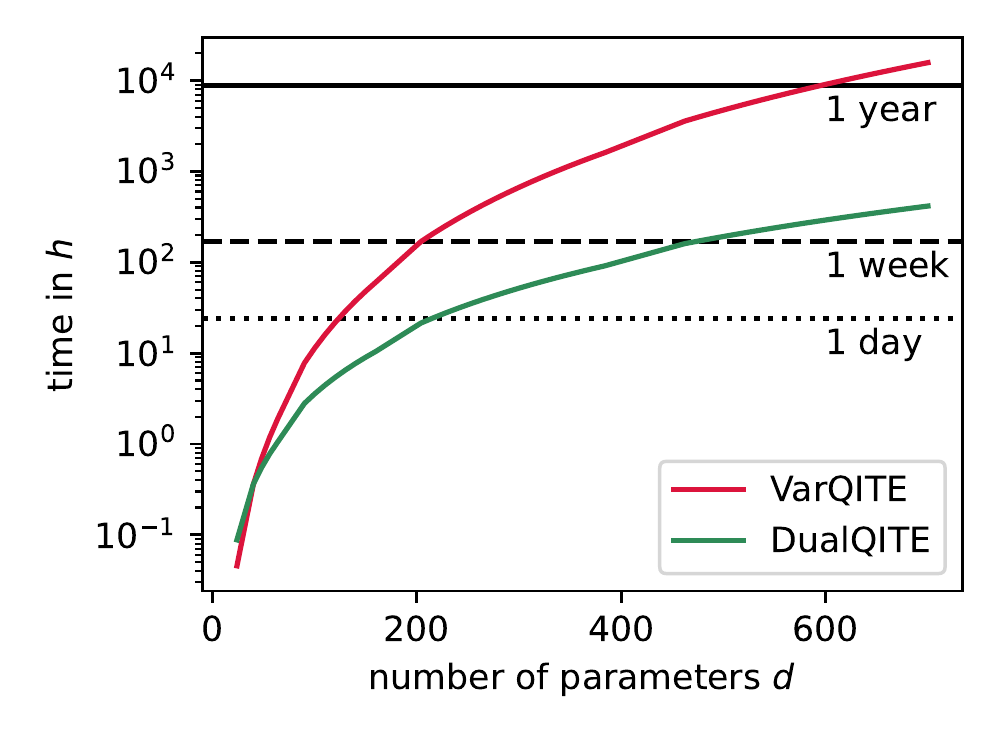}
    \caption{Estimated runtimes of variational imaginary-time evolution (VarQITE) and our proposed dual method (DualQITE) as a function of the number of parameters $d$ of the variational model, for an exemplary Heisenberg model and 200 timesteps.
    See Appendix~\ref{app:runtime_estimates} for more details.
    }
    \label{fig:varqte_scaling}
\end{figure}

Since performing quantum time evolution generally requires representing the exponentially large wave function of a quantum system, quantum computers are a promising platform for developing efficient algorithms~\cite{miessen_dynamics_2023}. 
In fact, in 1996, the Trotter algorithm for real-time evolution was among the first proposed use cases for a quantum computer~\cite{lloyd_universal_1996}. 
However, the complexity of the quantum circuits required for the Trotter algorithm depends on the Hamiltonian, and the circuit depth scales with the simulation time and accuracy~\cite{zhao_adatrotter_2022, mukhopadhyay_trotter_2023}. This renders the algorithm currently unsuitable for general time evolution on near-term devices, which are characterized by limited qubit connectivity and coherence times. The imaginary-time counterpart of Trotter suffers from the same restriction~\cite{motta_determining_2020}.

Variational algorithms for quantum time evolution, on the other hand, allow to choose a parameterized circuit as an ansatz to approximate the wave function, that operates within the device's capabilities.
Using a variational principle, variational quantum time evolution (VarQTE) maps the quantum state evolution to the evolution of parameters in the model \cite{yuan_varqte_2019}, both for \emph{real}-time evolution (VarQRTE) and \emph{imaginary}-time evolution (VarQITE).
The parameter update rules depend on the evaluation of the Quantum Geometric Tensor (QGT) and gradients of the current energy and state. For an ansatz with $d$ variational parameters, the number of circuits required to evaluate the QGT and gradients scale as $\mathcal{O}(d^2)$ and $\mathcal{O}(d)$, respectively. 
While this does not pose a problem for small systems, the evaluation of the QGT quickly becomes a bottleneck once the system size, and therefore the number of variational parameters, increases.

Figure~\ref{fig:varqte_scaling} shows a runtime estimation for VarQITE, assuming current superconducting processor specifications, see Appendix~\ref{app:runtime_estimates} for details on the derivation. 
For a few parameters the runtime is of the order of hours, but already for only 200 parameters the computation time around 1 week, which renders this algorithm currently impractical.
With recent advances in processor sizes exceeding 100 qubits, such as the IBM Quantum Eagle or Osprey devices~\cite{eagle, osprey}, improving the resource requirements of quantum algorithms becomes crucial for finding practically relevant applications of quantum computers.

Recently, focus has shifted to optimization-based algorithms, which implement partial steps or approximations of the full Suzuki-Trotter step~\cite{barison_pvqd_2021, barratt_parallel_2021, lin_compressed_2021, benedetti_evolution_2021, slattery_ublock_2022}. 
In the case of real-time evolution, for example, the projected variational quantum dynamics (p-VQD) algorithm~\cite{barison_pvqd_2021} provides a scalable alternative to VarQRTE on near term-devices, if a single Trotter step can be efficiently implemented.
However, the required quantum circuit gates in p-VQD reflect the couplings of the Hamiltonian. This means that, for Hamiltonians with long-distance interactions or numerous Pauli terms (e.g. in molecular dynamics), even a single step could involve global connections or deep circuits hindering the execution on near-term devices.
Furthermore, the p-VQD algorithm is not directly applicable to imaginary-time evolution.

Other approaches concerned with real-time evolution are Variational Fast Forwarding (VFF) methods \cite{commeau_vffhamiltonian_2020, cirstoiu_vfftrotter_2020, gibbs_vffdemo_2021, lim_vffdemo_2022} and classical pre-processing approaches \cite{mckeever_classicalopt_2023, mansuroglu_classicalpre_2023}.
VFF methods rely on diagonalizing the Hamiltonian or the Trotterized time evolution operator with a variational ansatz. However, finding the diagonalizing unitary remains challenging in practice, which limits demonstrations to very few qubits. 
Classical pre-processing techniques, on the other hand, impose additional restrictions on the simulated system, such as translational invariance \cite{mansuroglu_classicalpre_2023} or Hamiltonians with low entanglement \cite{mckeever_classicalopt_2023}. Within such systems, these techniques scale to large systems, but they do not allow for general quantum time evolution.

Another line of work directly focuses on the preparation of thermal states by minimizing the free energy of a variational ansatz \cite{sbahi_mirror_2022}. This approach, however, also does not implement general quantum time evolution.

In this paper, we propose a novel variational algorithm for quantum time evolution based on a \emph{dual} optimization problem, which allows to replace the QGT by evaluating the overlap of the variational ansatz for different parameter values. This formulation applies equally real- and imaginary-time evolution and does not require additional qubits or connections than already present in the ansatz. We show that this new algorithm requires significantly fewer measurements and thereby drastically reduces the expected runtime compared to VarQTE. This is summarized in Fig.~\ref{fig:varqte_scaling}, where, under the same assumptions, our proposed method can reduce the expected runtimes from several weeks for VarQTE to only a few days. Following the naming conventions of VarQTE, we name the algorithm DualQTE with specifiers DualQITE for imaginary-time evolution and DualQRTE for real-time evolution.

The remainder of this paper is structured as follows.
In Sec.~\ref{sec:theory}, we recap VarQTE based on variational principles, derive the proposed dual formulation, and discuss how to implement it on a quantum computer. Then, in Sec.~\ref{sec:imaginary}, we demonstrate our proposed algorithm for the imaginary-time evolution of the Heisenberg model and investigate the resource requirements. As a practical application, we use the quantum minimally entangled typical thermal states method (QMETTS) to calculate thermodynamic observables. 
Sec.~\ref{sec:real} demonstrates the dual formulation for real-time evolution, including the calculation of variational error bounds.
Finally, Sec.~\ref{sec:conclusion} concludes the paper and gives an outlook on possible applications and further research directions.

\section{Dual formulation of variational time evolution}
\label{sec:theory}

For a time-independent Hamiltonian $H$ acting on $n$ qubits, an initial quantum state $\ket{\Psi_0}$ and an evolution time $t$, the real-time evolved quantum state is
\begin{equation}
    \ket{\Psi(t)} = e^{-it H}\ket{\Psi_0}.
\end{equation}
For an imaginary-time evolution, the time evolution operator is non-unitary, and the normalized  state reads
\begin{equation}
    \ket{\Psi(t)} = \frac{1}{\sqrt{\braket{\Psi_0 | e^{-2tH} | \Psi_0}}} e^{-tH}\ket{\Psi_0}.
\end{equation}

Variational quantum time evolution maps the evolution of the quantum state $\ket{\Psi(t)}$ to the evolution of parameters $\vec\theta(t) \in \mathbb{R}^d$ of a parameterized quantum state 
$\ket{\phi(\vec\theta(t))}$. The parameters' dynamics can be derived with variational principles such as the Dirac-Frenkel, McLachlan, or time-dependent variational principle \cite{yuan_varqte_2019}.
In McLachlan's formulation, the derivatives of the parameters are determined by the linear system of equations
\begin{equation}\label{eq:lse}
    g(\vec\theta(t))\,\dot{\vec\theta}(t) = \vec b(\vec\theta(t)),
\end{equation}
where the matrix $g = \mathrm{Re}(G) \in \mathbb{R}^{d\times d}$ is the real part of the QGT, and we call $b \in \mathbb{R}^d$ the evolution gradient. 

The QGT is defined as
\begin{equation}\label{eq:qgt}
    G_{ij}(\vec\theta) = \braket{\partial_i \phi(\vec\theta)|\partial_j \phi(\vec\theta)} - \braket{\partial_i \phi(\vec\theta)|\phi(\vec\theta)}\braket{\phi(\vec\theta)|\partial_j \phi(\vec\theta)},
\end{equation}
where we use the notation $\partial_i := \partial/(\partial \theta_i)$ and do not explicitly state the time dependence of the parameters.
The evolution gradient for VarQRTE is given by the expression
\begin{equation}\label{eq:b_real}
    b^\text{R}_i(\vec\theta) = \mathrm{Im}\big(\braket{\partial_i\phi(\vec\theta)|H|\phi(\vec\theta)} -\braket{\partial_i\phi(\vec\theta)|\phi(\vec\theta)} E(\vec\theta) \big), 
\end{equation}
whereas a VarQITE evolution yields the following 
\begin{equation}\label{eq:b_imag}
    b^\text{I}_i(\vec\theta) = -\mathrm{Re}\big(\braket{\partial_i\phi(\vec\theta)|H|\phi(\vec\theta)}\big) = -\frac{\partial_i E(\vec\theta)}{2},
\end{equation}
with the energy $E(\vec\theta) = \braket{\phi(\vec\theta)|H|\phi(\vec\theta)}$. 
From hereon, we present general equations that apply to both real and imaginary-time evolution; thus, unless specified, we simply use $b$ without a specific superscript.

Note that these equations are introduced for a time-independent Hamiltonian, but they can also be applied to the time-dependent case $H = H(t)$.

\subsection{Dual formulation}

Instead of solving the linear system defined in Eq.~\eqref{eq:lse}, we propose to solve the dual formulation of the problem ~\cite{amari_natural_1998, stokes_qng_2020} given by
\begin{equation}\label{eq:argmin_qgt}
   \dot{\vec\theta} = \argmin{\dot{\vec\theta}} \frac{\dot{\vec\theta}^T g(\vec\theta) \dot{\vec\theta}}{2} - \dot{\vec\theta}^T \vec{b}(\vec\theta).
\end{equation}
The term $\|\dot{\vec\theta}\|^2_{g(\vec\theta)} = \dot{\vec\theta}^T g(\vec\theta) \dot{\vec\theta}$ is the squared norm of the parameter derivative in the metric of the QGT.
This quantity describes the magnitude of the derivative from an information geometric point of view and is derived from the Fubini-Study metric.
For infinitesimal displacements $\vec{\delta\theta}$, we have 
\begin{equation}\label{eq:qgt_approximation}
    \begin{aligned}
        ||\vec{\delta\theta}||^2_{g(\vec\theta)} &= \vec{\delta\theta}^T g(\vec\theta) \vec{\delta\theta} \\
                                       &= 1 - |\braket{\phi(\vec\theta)|\phi(\vec\theta + \vec{\delta\theta})}|^2 + \mathcal{O}(\|\vec{\delta\theta}\|_2^3),
    \end{aligned}
\end{equation}
where $\|\cdot\|_2$ is the $\ell_2$ norm~\cite{stokes_qng_2020}.
By writing $\dot{\vec\theta} = \vec{\delta\theta} / \delta\tau$, for some time perturbation $\delta\tau > 0$, we can now reformulate the optimization in terms of the fidelity $F(\vec{\theta}, \vec{\theta'}) = |\braket{\phi(\vec\theta)|\phi(\vec\theta')}|^2$ as
\begin{equation}\label{eq:argmin_dual}
    \begin{aligned}
   \vec{\delta\theta}
   &\approx \argmin{\vec{\delta\theta}} \frac{1 - F(\vec\theta, \vec\theta+\vec{\delta\theta})}{2(\delta\tau)^2} - \frac{\vec{\delta\theta}^T \vec b(\vec\theta)}{\delta\tau} \\
   &= \argmin{\vec{\delta\theta}} \frac{\mathcal{L}(\vec{\delta\theta})}{(\delta\tau)^2},
   \end{aligned}
\end{equation}
where we directly optimize for the parameter update $\vec{\delta\theta}$ and we introduced the loss function 
\begin{equation}\label{eq:dual_cost}
\mathcal{L}(\vec{\delta\theta}) = \frac{1 - F(\vec\theta, \vec\theta+\vec{\delta\theta})}{2} - \delta\tau \cdot \vec{\delta\theta}^T \vec b(\vec\theta).
\end{equation}
In practice, the optimization problem can be solved without the factor $(\delta\tau)^{-2}$, which decouples the shape of the locally quadratic infidelity term from the time perturbation and improves the numerical stability of the optimization.

Note that this dual formulation can alternatively be obtained from the derivation of quantum natural gradients \cite{stokes_qng_2020, sbahi_mirror_2022}, which is detailed in Appendix~\ref{app:qng_derivation}.
For an intuitive understanding of the relationship of the infidelity and QGT the effect of approximating $\|\vec{\delta\theta}\|_{g(\vec\theta)} \approx 1 - F(\vec\theta, \vec\theta + \vec{\delta\theta})$ in an illustrative example is demonstrated in Appendix~\ref{app:illustrative_example}.

Instead of computing the QGT at each timestep, which requires $\mathcal{O}(d^2)$ circuit evaluations, we now have to solve an optimization problem where the loss function requires only one fidelity evaluation. 
The required resources of DualQTE per timestep are therefore $\mathcal{O}(d)$ for the computation of the evolution gradient $b$, times the number of iterations in the optimization. 
Thus, we improve upon the direct QGT approach if the number of iterations scales better than $\mathcal{O}(d)$, which, as we show in the following sections, is the case for the examples we investigate in this work.

\subsection{Evaluating the loss function}

The evaluation of the loss function $\mathcal{L}$, defined in Eq.~\eqref{eq:dual_cost}, requires the calculation of the evolution gradient $b$ and the fidelity of the ansatz $\ket{\phi(\vec\theta)}$ for two different parameter sets. 
For imaginary-time evolution, the evolution gradient can, for example, be evaluated with analytic gradient rules, such as the parameter-shift rule or a linear combination of unitaries (LCU), or with finite difference methods~\cite{schuld_gradients_2019}.
In the case of real-time evolution, however, we are restricted to an LCU approach, as this is the only method that allows the calculation of the imaginary part of gradients~\cite{zoufal_variational_2021}. 

The fidelity $F$ can, for example, be estimated using the swap test~\cite{buhrman_swaptest_2001} and its variants~\cite{cincio_learning_2018}, where the states are prepared in separate qubit registers followed by entangling gates across these registers, or with the Hadamard test, which adds only a single auxiliary qubit, but requires controlling the state-preparing unitary \cite{cleve_hadamardtest_1998}.
A more near-term-friendly option is the compute-uncompute method~\cite{havlicek_supervised_2019}, which does not introduce additional global operations.
If the states are given by $\ket{\phi(\vec\theta)} = U(\vec\theta)\ket{0}$ for a parameterized unitary $U$ and two different parameter values $\vec\theta$ and $\vec\theta'$, the fidelity can be calculated by preparing $U^\dagger(\vec\theta) U(\vec\theta')\ket{0}$ and measuring the probability of obtaining $\ket{0}$ on all qubits. 

If the state $\ket{\phi(\vec\theta)}$ has $n$ qubits and the preparing unitary $U$ has depth $m$, the swap test variants require a circuit width of 
$2n$ with depth of $m + \mathcal{O}(1)$, whereas the evaluated circuits for the compute-uncompute method are of only width $n$, but of depth $2m$.
The Hadamard test for fidelities between the same circuit with different parameters can be evaluated by controlling the parameterized gates, resulting in depth of $m$ and depth of $n + 1$, plus the overhead of controlling the gates. For sparse device connectivities, this can be a challenge.
To avoid increasing the circuit complexity, the overlap can also be estimated via randomized measurements of two independent state preparations~\cite{elben_overlap_2019}. However, this technique requires an exponential number of measurements.

Evaluating the QGT for VarQTE, however, suffers from similar issues. The QGT can be evaluated as the Hessian of the infidelity~\cite{gacon_qnspsa_2021} using a parameter-shift or finite difference technique, which comes with the restrictions for fidelity evaluations as described above. Alternatively, Eq.~\eqref{eq:qgt} can be directly computed with an LCU approach, which adds two auxiliary qubits and two entangling gates~\cite{zoufal_variational_2021}. This method is less demanding than e.g. a Hadamard test, but still comes with additional connectivity requirements.
In practice, for both VarQTE and DualQTE a suitable combination of parameterized quantum state $\ket{\phi(\vec\theta)}$ and gradient method must be selected, such that the resulting circuits can be executed reliably.

Depending on the topology and coherence times of the available hardware and the structure and size of the unitary, either method for gradient and fidelity calculations can be advantageous.
In this work, we focus on near-term friendly methods and use the parameter-shift rule for gradients (if possible) and the compute-uncompute method for the fidelity, as these do not require additional gate connections or an exponential number of measurements. 
Note that, for systems with a large number of qubits, this method might become unsuitable as it measures the global zero projector. Then, approaches using only local measurements, such as the Hadamard test, could be the better choice.

\subsection{Solving for the update step}

The infidelity-based loss function $\mathcal{L}$ is locally convex around $\vec{\delta\theta} = \vec{0}$, as its Hessian at this point is $\vec\nabla\vec\nabla^T \mathcal{L}(0) = g/2$, and $g$ is positive semi-definite. To leverage this property, we use gradient descent as a local optimization routine, which also allows the use of analytic gradient formulas that have proven more stable in presence of shot noise.
The gradient of $\mathcal{L}$ with respect to the parameter update $\vec{\delta\theta}$ is 
\begin{equation*}
    \vec\nabla_{\vec{\delta\theta}} \mathcal{L}(\vec{\delta\theta}) = -\frac{\vec\nabla_{\vec{\delta\theta}} F(\vec\theta, \vec\theta + \vec{\delta\theta})}{2} - \delta\tau \cdot \vec{b}(\vec\theta).
\end{equation*}
The gradient of the fidelity can be evaluated with a parameter-shift rule
\begin{equation*}
    \frac{\partial F}{\partial (\delta\theta)_i} = \frac{F(\vec\theta, \vec\theta + \vec{\delta\theta} + \vec e_i s) - F(\vec\theta, \vec\theta + \vec{\delta\theta} - \vec e_i s)}{2\sin(s)},
\end{equation*}
where $\vec e_i$ is the $i$-th unit vector and $s$ is the parameter shift, which can be chosen as, e.g., $\pi/2$ for single-qubit Pauli rotations \cite{schuld_gradients_2019}.

At each timestep, the gradient descent update for the update step $\vec{\delta\theta}$ is
\begin{equation*}
    \vec{\delta\theta}^{(k+1)} = \vec{\delta\theta}^{(k)} - \eta_k \vec\nabla \mathcal{L}\left(\vec{\delta\theta}^{(k)}\right),
\end{equation*}
where $\eta_k > 0$ is the learning rate at step $k$.
This iteration is continued until a maximum number of iterations or a convergence criterion is met. An example of the latter is a minimum tolerance in the change of the cost function or the norm of the gradient.

An intuitive choice for the initial guess $\vec{\delta\theta}^{(0)}$ is the zero vector, which corresponds to no change in the parameters. However, a more efficient choice can be to introduce momentum by warm starting the optimization with the update step from the previous timestep. This heuristic is motivated by the idea that, especially for small timesteps, we do not expect the parameter derivatives $\dot{\vec\theta}$ to change significantly. 

Methods that approximate the gradient, such as finite difference or SPSA~\cite{spall_spsa_1998}, may face challenges in the optimization.
For small timesteps, the fidelity is close to $1$ and the noise in the readout, e.g. from finite sampling statistics or device noise, can easily mask changes in the cost function. Parameter-shift gradients suffer less from this problem, as they allow to evaluate the cost function over larger perturbations, and do not amplify the noise by dividing by a small constant.

The ideal choice of the time perturbation $\delta\tau$ is a trade-off: the error in approximating the QGT scales with $(\delta\tau)^3$, but a smaller perturbation amplifies any measurement noise in the loss function as the update step is obtained as $\vec{\delta\theta} / \delta\tau$. Appendix~\ref{app:illustrative_example} displays this trade-off for an illustrative example.

\subsection{Trainability}

Recently, there has been a lot of research showing that, in certain settings, the loss function gradients of variational algorithms decay to zero exponentially and cannot be evaluated efficiently, as they would require an exponential number of measurements. These so-called barren plateaus can be induced, for example, if the loss function requires measuring a global observable \cite{cerezo_cost-induced_2021}, if the quantum circuit preparing the parameterized state is too deep or generates too much entanglement \cite{mcclean_barren_2018, cerezo_cost-induced_2021, ortiz_entanglement-induced_2021}, or if the measurements are too noisy \cite{wang_noise-induced_2021}.

Since variational quantum dynamics is driven by the evolution gradient defined Eqs.~\eqref{eq:b_real} and~\eqref{eq:b_imag}, it can be affected by a barren plateau and fail to track the true evolution of the quantum state. 
However, it is important to note that the gradients only vanish on average for a random initialization, whereas in time-evolution the initial quantum state is typically specifically chosen.
Furthermore, Hamiltonians of physical systems are usually local, as they reflect the interactions of the quantum mechanical system, and exponentially vanishing gradients can be avoided by choosing a circuit depth scaling logarithmically in system size \cite{cerezo_cost-induced_2021}. Alternatively, an application-specific ansatz with few variational parameters can help mitigate barren plateaus, such as circuits based on Hamiltonian evolutions \cite{ollitrault_uvcc_2020, park_hva_2023}.

In addition to the evolution gradient, the DualQTE loss function gradient $\vec\nabla_{\vec{\delta\theta}} \mathcal{L}$ depends on the gradient of the fidelity, which relies on measuring a global observable. This can be seen by writing the fidelity of two $n$-qubit states prepared by unitaries $U(\vec\theta)$ and $U(\vec\theta')$ as 
$|\braket{0|U^\dagger(\vec\theta) U(\vec\theta')|0}|^2 = \braket{\lambda|P_0|\lambda}$, where $\ket{\lambda} = U^\dagger(\vec\theta') U(\vec\theta)\ket{0}$ and $P_0 = \ket{0}\bra{0}^{\otimes n}$ is the global projector on the all-zero state. 
Thus, evaluating the fidelity gradient for two randomly selected parameter sets $\vec\theta$ and $\vec\theta'$ would exhibit barren plateaus at any circuit depth \cite{cerezo_cost-induced_2021}.
However, the optimization in DualQTE starts at zero perturbations, $\vec\theta = \vec\theta'$ where the total state preparing unitary is the identity, $\ket{\lambda} = U^\dagger(\vec\theta)U(\vec\theta)\ket{0} = \mathbb{I}\ket{0}$, which is an initialization that is proven to not exhibit barren plateaus  even for global cost functions~\cite{grant_bp-initialization_2019}.
Together with the fact that the DualQTE loss function is locally convex, 
the non-vanishing gradients at the initial point of the optimization is a strong motivation for the efficient trainability of DualQTE.

In Appendix~\ref{app:vanishing_gradients}, we provide numerical evidence that for a local Hamiltonian and a logarithmic-depth circuit, neither the evolution gradient or the fidelity gradients decay exponentially with system size.

\subsection{Sample complexity}\label{sec:sampling}

The implementation of VarQTE on quantum hardware has several sources of errors:
the model $\ket{\phi(\vec\theta)}$ could lack expressitivity to capture the dynamics, the time integration scheme introduces errors, the QGT and evolution gradient are subject to sampling error from a finite number of measurement, and each operation is affected by hardware noise.
If we denote the ideal VarQTE parameters without sampling or hardware noise by $\vec\theta(t)$ and the noisy parameters by $\tilde{\vec\theta}(t)$, the error contributions can be split as
\begin{equation}
    \begin{aligned}
        \varepsilon(t) &= D_B(\phi(\tilde{\vec\theta}(t)), \Psi(t)) \\
        &\leq \varepsilon_M(t) + \varepsilon_S(t),
    \end{aligned}
\end{equation}
where we measured the error in Bures distance
\begin{equation}
    D_B(\psi, \phi) = \sqrt{2(1 - |\braket{\psi|\phi}|)},
\end{equation}
and distinguish in error due to lack of model expressitivity plus integration error $\varepsilon_M(t) = D_B(\phi(\vec\theta(t)), \Psi(t))$ and
error due to a noisy implementation of VarQTE $\varepsilon_S(t) =  D_B(\phi(\tilde{\vec\theta}(t)), \phi(\vec\theta(t))$ \cite{endo_varqte_2020}.

Since the proposed DualQTE algorithm promises a reduction in measurement cost of VarQTE, but is not concerned with the ansatz selection or hardware noise, we here focus on investigating the scaling of the sampling error.
The model error $\varepsilon_M$ can be bounded with a-posteriori errorbounds \cite{zoufal_errorbounds_2021}, which we also investigate for the real-time evolution case in Sec.~\ref{sec:real}.

Deriving a concrete bound in terms of system quantities such as the energy or the number of parameters requires an assumption on the circuit structure. Here we assume a circuit with only Pauli rotations $R_P(\theta_i)$ where each parameter $\theta_i$ is unique and does not have coefficients. Note that the bounds can be adjusted for different circuit structures and parameterizations.
In addition, we assume a cutoff $\delta_c > 0$ on the smallest eigenvalue of $g$ due to a regularization of the linear system. 

Then, we can state the following upper bound on the number of samples required to achieve a sampling error of $\varepsilon_S$,
\begin{equation}\label{eq:vq_upper_bound}
    N \leq \mathcal{O}\left(\frac{d^3 E_\text{max}^2 \Delta_t^2}{\delta_c^4 \varepsilon_S^2}\right),
\end{equation}
where $E_\text{max}$ is the maximal eigenvalue of the Hamiltonian. 

In contrast to a similar approach described in Ref.~\cite{endo_varqte_2020}, we state the upper bound in terms of the number of parameters in the model or the system's Hamiltonian, instead of the QGT and evolution gradients. Further, we are able to derive a tighter result by leveraging Latala's theorem from random matrix theory to upper bound the sampling error in $g$ \cite{latala_bound_2005}.

Since the DualQTE algorithm does not construct the QGT directly but only the evolution gradient, we expect a reduction of a factor $d$ in the complexity, and an additional factor for the number of optimization steps $K$ in each timestep.
Indeed we can show that the upper bound for the number of samples is
\begin{equation}\label{eq:dual_upper_bound}
    N \leq \mathcal{O}\left(\frac{d^2 K^2 \Delta_t^2}{\delta\tau^2 \varepsilon_S^2}\left(\frac{1}{\delta\tau} + E_\text{max}\right)^2\right).
\end{equation}
The detailed derivation of both bounds is described in Appendix~\ref{app:sampling_error}.

While it is possible to construct circuits where each component of these bounds are tight Sec.~\ref{sec:imaginary} shows that in practice the actual number of required samples scales less than this upper bound, which is further discussed in the Appendix.

\begin{figure*}[htp]
    \centering
    \includegraphics[width=0.49\linewidth]{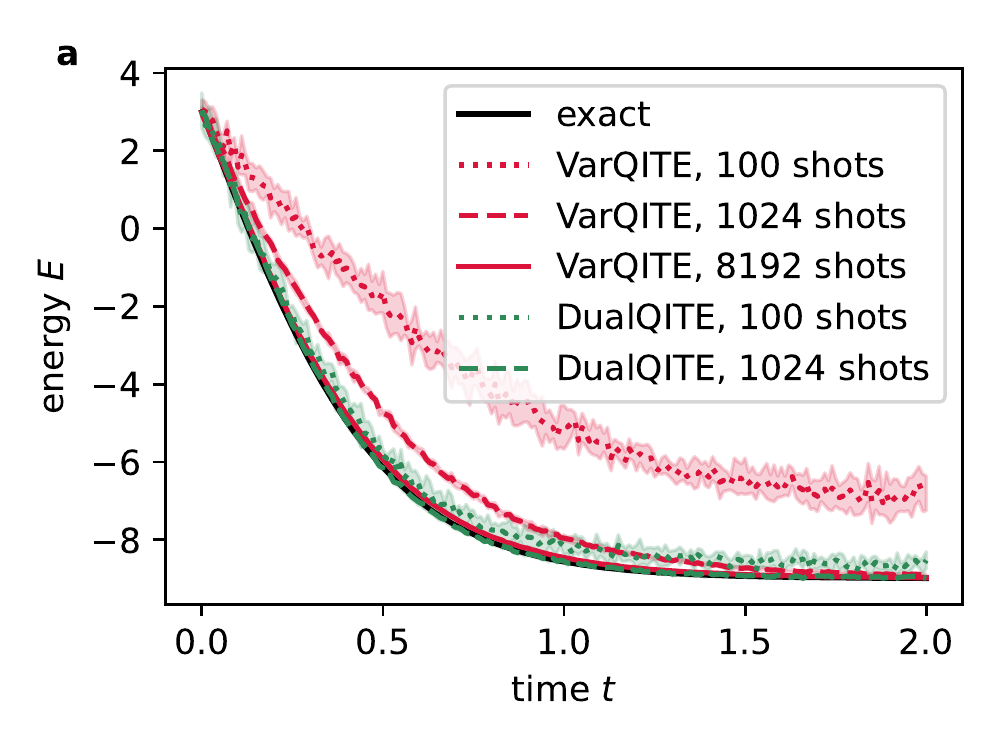}
    \includegraphics[width=0.49\linewidth]{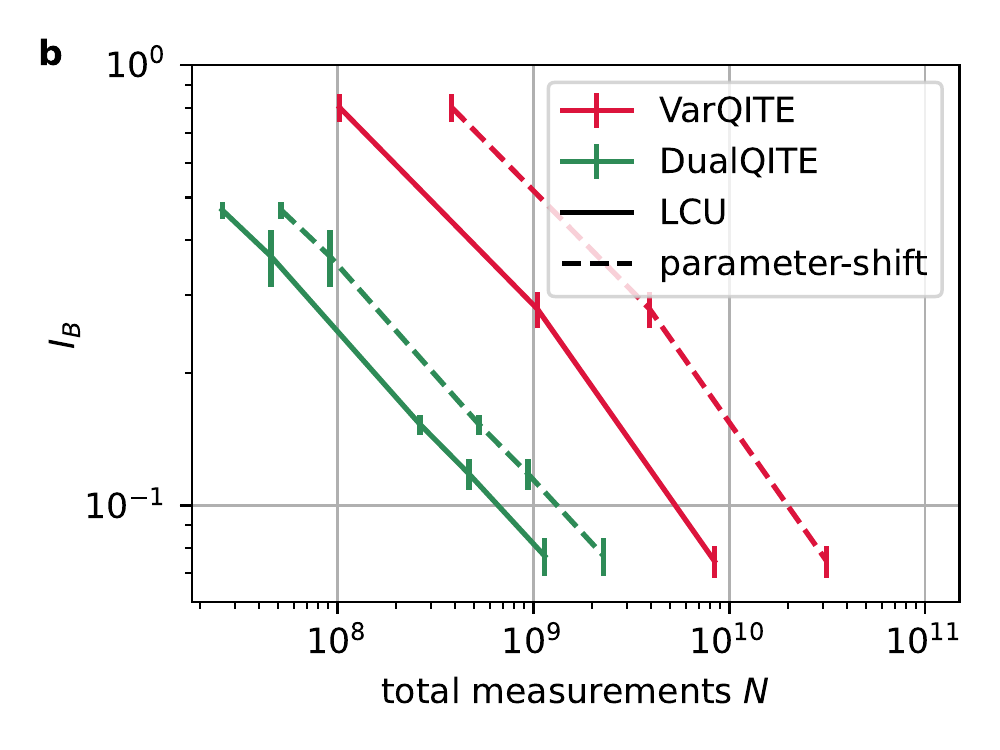}
    \caption{
    (a) The mean and standard deviation of DualQITE and VarQITE, each averaged over 5 independent experiments for a varying number of shots. 
    (b) The accuracy measured in integrated Bures distance $I_B$ (see Eq.~\eqref{eq:integrated_bures}) for DualQITE and VarQITE, with mean and standard deviation of 5 experiments. The resources are measured in total number of measurements and are shown for evaluation of gradients (and the QGT, in the case of VarQITE) using parameter-shift rules (dashed) or a LCU approach (solid lines).
    }
    \label{fig:heisen}
\end{figure*}

\section{Imaginary-time evolution}
\label{sec:imaginary}

In this section, we show the results of DualQITE and investigate the circuit costs compared to VarQITE.
As an application, we use our algorithm as a subroutine to prepare typical thermal states of the Heisenberg model, which are then used to calculate the energy per site as a thermodynamic observable.
All circuits are constructed and simulated using Qiskit~\cite{qiskit}.

\subsection{Heisenberg model}

We simulate the imaginary-time evolution of the Heisenberg model with nearest-neighbor interaction on a 12-qubit circle in  a transverse field:
\begin{equation}\label{eq:heisencomb}
    H = J \sum_{\braket{ij}} \left( X_i X_j + Y_i Y_j + Z_i Z_j \right) + g \sum_i Z_i,
\end{equation}
with interaction strength $J=1/4$ and transversal field strength $g=-1$.
As a variational ansatz, we use a circuit with Pauli-$Y$ and Pauli-$Z$ single qubit rotation layers that alternate with pairwise CNOT entangling gates. The circuit structure is 
shown in Appendix~\ref{app:su2_circuit} and we use $r=3$ repetitions. The initial state for the evolution is the equal superposition of all qubits, $\ket{+}^{\otimes n}$,
which we prepare by setting the rotation angles of the last Pauli-$Y$ layer to $\pi/2$ and the remaining angles to $0$.

The optimization problems in DualQITE are solved using gradient descent with a fixed learning rate of $\eta=0.1$ and time perturbation $\delta\tau = 0.01$. The initial iteration performs $100$ update steps and the subsequent, warmstarted iterations, only $10$. These values are motivated by the simulations shown in the Appendix~\ref{app:warmstarting} and are partially heuristic, as a termination criterion is challenging to define with access only to noisy loss function and gradient evaluations. 
The parameters are integrated with an explicit Euler scheme with timestep $\Delta_t = 0.01$, i.e.
\begin{equation*}
    \vec\theta(t + \Delta_t) = \vec\theta(t) + \Delta_t\dot{\vec\theta}(t) = \vec\theta(t) + \Delta_t\frac{\vec{\delta\theta}}{\delta\tau}.
\end{equation*}
Note that the integration timestep $\Delta_t$, which determines the accuracy of the time integration, can be chosen differently from the time perturbation $\delta\tau$, which affects the approximation error of the QGT metric with the infidelity.

We compare the performance to VarQITE with the same integration scheme and use an L-curve regularization~\cite{cultrera_lcurve_2020} 
for a stable solution of the linear system. Among all regularization techniques we attempted, such as adding a diagonal shift,
truncating small or negative singular values or solving on a stable subsystem, the L-curve regularization provided the most accurate and stable results.

In Fig.~\ref{fig:heisen}(a), we present the results for a varying number of shots along with the exact time evolution based
on exact diagonalization. Already with as little as $100$ measurements per circuit evaluation (shots) on the $12$-qubit model, the dual time evolution is able 
to qualitatively follow the imaginary-time evolution and, up to time $t \approx 1$, even outperform VarQITE with $1024$ shots.
Increasing the number of measurements of DualQITE to $1024$ shots allows the dual method to closely track the exact solution towards the ground state, with a higher accuracy than VarQITE with $8192$ shots. 

\subsection{Resource requirements}

In the above experiment, DualQITE requires fewer circuit evaluations to achieve the same accuracy as VarQITE. To investigate the total resource requirements, 
we perform both DualQITE and VarQITE with different resources and compute the achieved error.
Since we are interested in following the imaginary-time dynamics as closely as possible at each timestep, we define the error as the average integrated Bures distance to the exact solution over the time evolution,
\begin{equation}\label{eq:integrated_bures}
    I_B(T) = \frac{1}{T}\int_0^T D_B(\phi(\vec\theta(t)), \psi(t)) \mathrm{d}t.
\end{equation}
The state fidelity is computed exactly, i.e., we compute the state vector of the model $\ket{\phi}$ at variational parameters $\vec\theta(t)$, 
and take the inner product with the exact time-evolved state $\ket{\Psi(t)}$. 

The results for an integration time of $T=2$ are shown in Fig.~\ref{fig:heisen}(b). We show the integrated Bures distance with respect to the total number of measurements recorded during the time evolution.
In DualQITE, the resources can be split between using more optimization steps in each timestep or more shots to evaluate the gradients. The algorithm settings are detailed in Appendix~\ref{app:dual_resources}. The figure shows the resource counts for gradient calculations via the parameter-shift rule (PSR) and linear combination of unitaries (LCU). The LCU technique requires additional auxiliary qubits and additional non-local operations, but less overall circuits than PSR. For $P$ Pauli terms in the Hamiltonian, the total number of required circuits $C$ per timestep
\begin{equation}
    \begin{aligned}
        C^\text{VarQITE}_\text{LCU} &= \frac{d(d+5)}{2} + Pd  \\
        C^\text{VarQITE}_\text{PSR} &= 2d(d + P + 1).
    \end{aligned}
\end{equation}
For DualQTE the number of circuits is
\begin{equation}
    C^\text{DualQITE}_\text{LCU} = Pd + Kd,
\end{equation}
and $C^\text{DualQITE}_\text{PSR} = 2 C^\text{DualQITE}_\text{LCU}$, where $K$ is the number of optimization steps per timestep.
The total number of measurements $N$ is obtained by multiplying the number of circuits with the number of shots.

We see that, on average, DualQITE requires about one order of magnitude fewer measurements to achieve the same accuracy as VarQITE. 
With an increasing number of parameters, we expect this difference to grow, since VarQITE scales as $\mathcal{O}(d^2)$ whereas our algorithm, with warm starting, only computes small corrections at each time step.

\subsection{Sample complexity}\label{sec:complexity_benchmark}

In addition to the fixed-size model with 12 qubits, we investigate how the resource requirements scale with system size. 
We compare VarQITE and DualQITE for the Heisenberg model from Eq.~\eqref{eq:heisencomb} with varying number of spins $n$
and the same circuit structure as before, but with an adjusted number of repetitions of $r=\lceil\log_2(n)\rceil$ times, plus a final rotation layer.
We then tune the settings of VarQITE and DualQITE to achieve a mean accuracy of $I_B \leq 0.1$ over 5 experiments and count the total number of required measurements $N$. This threshold corresponds to a per-timestep fidelity of 0.995. 

\begin{figure}[htp]
    \centering
    \includegraphics[width=\linewidth]{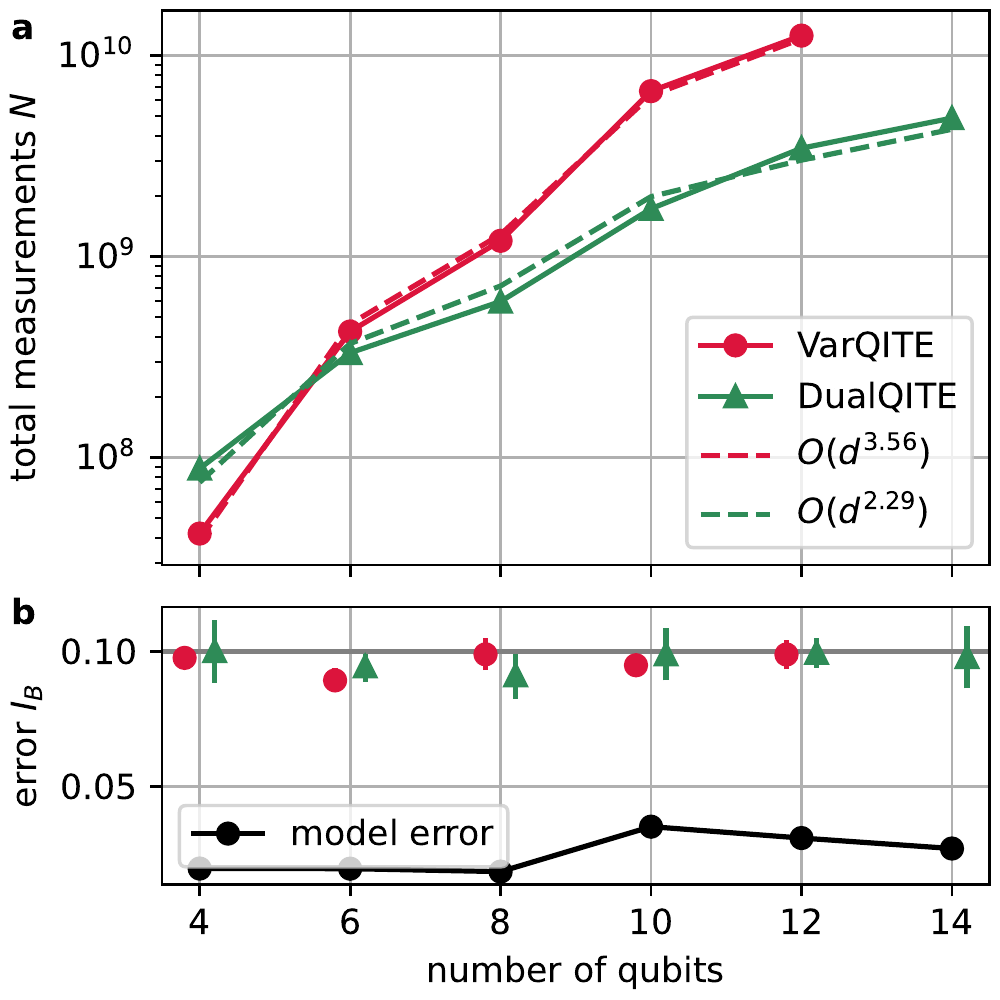}
    \caption{(a) Total number of measurements required to achieve a mean accuracy of $I_B \leq 0.1$ over an average of 5 experiments.
    See Table~\ref{tab:resources_sizescaling} for the exact algorithm settings. Dotted and dashed lines show fits for the number of measurements.
    The bumps in the fits are due to the discontinuity of the circuit depth, which depends on $\lceil\log_2(n)\rceil$. VarQITE is not evaluated
    for $n=14$ qubits as it requires too many measurements.
    (b) Mean accuracy and standard deviation of each point of the top panel. The grey line indicates the infidelity threshold of $I_B = 0.1$.}
    \label{fig:sizescaling}
\end{figure}

The results are presented in Fig.~\ref{fig:sizescaling}, which show the improved scaling of DualQITE compared to VarQITE. For small system sizes and few parameters, the overhead of solving the optimization problem in DualQITE is larger than evaluating the QGT. But, as we increase the problem size, the quadratic scaling of VarQITE takes over and our algorithm becomes more efficient.

This experiment allows to validate the upper bound on the number of measurements of Sec.~\ref{sec:sampling}. 
As shown in Fig.~\ref{fig:sizescaling}, the model error $\varepsilon_M$ is negligible in comparison to the sampling error $\varepsilon_S$ and we approximately have $\varepsilon_S \approx I_B \approx 0.1$.
The maximal energy of the Heisenberg model on a periodic chain scales with the number of spins, and can be bounded by $E_\text{max} = \mathcal{O}(n) \leq \mathcal{O}(n\log(n)) = \mathcal{O}(d)$. Inserting these values in Eqs.~\eqref{eq:vq_upper_bound} and~\eqref{eq:dual_upper_bound} we expect the scaling to be upper bounded by $\mathcal{O}(d^5)$ for VarQTE and $\mathcal{O}(d^4 K)$ for DualQTE.
In practice, we observe approximately a scaling of $d^{3.56}$ for VarQTE and $d^{2.29}$ for DualQTE, which shows the expected improved scaling for our algorithm. 
The measured scaling also suggests that the bounds are not yet tight, which we discuss further in Appendix~\ref{app:sampling_error}.

\subsection{Calculating thermodynamic observables}

As an application of imaginary-time evolution, we calculate thermodynamic observables using the quantum minimally entangled thermal states algorithm (QMETTS) \cite{stoudenmire_minimally_2010,motta_determining_2020}.
While the classical METTS algorithm has been specifically developed for Matrix Product State simulations, the thermal state preparation is still costly, and classical simulations fail if the system produces macroscopic entanglement during the imaginary time evolution (e.g. for low-temperature, 2D systems). Due to these restrictions, QMETTS is a promising application for quantum imaginary-time evolution algorithms.

For an observable $A$ and inverse temperature $\beta$, the QMETTS algorithm generates samples $\{A_m\}_m$ using a Markov chain whose average approximate the 
ensemble average:
\begin{equation*}
    \braket{A}_\text{ens} = \frac{\tr(e^{-\beta H} A)}{\tr(e^{-\beta H})} \approx \frac{1}{M} \sum_{m=1}^{M} A_m.
\end{equation*}
The sampling process to obtain the sample $A_m$ is
\begin{enumerate}
    \item Start from a product state $\ket{\phi_m(t=0)}$.
    \item Evolve up to imaginary time $t = \beta/2$
    \begin{equation*}
        \ket{\phi_m(\beta/2)} \propto e^{-\beta H/2 }\ket{\phi_m(0)}.
    \end{equation*}
    \item Evaluate the observable to obtain the sample
    \begin{equation*}
        A_m = \braket{\phi_m(\beta/2) | A |\phi_m(\beta/2)}.
    \end{equation*}
    \item Measure $\ket{\phi_m(\beta/2)}$ in some basis to obtain
    the next random product state $\ket{\phi_{m+1}(0)}$.
\end{enumerate}

We investigate the Heisenberg model from Eq.~\eqref{eq:heisencomb} on a chain with $n=6$ spins
with parameters $J=1/4$ and $g=-1$. As a thermodynamic observable we 
compute the energy per site, $\braket{H}/n$.
To reduce the auto-correlation length in the QMETTS Markov chain, and for 
faster convergence to the ensemble average, it is favorable to measure in 
different bases in each step. 
Since the Heisenberg Hamiltonian conserves the number of qubits in the $\ket{1}$ state, avoiding the $Z$ basis greatly reduces the standard deviation of the Markov chain. Thus,
we here alternate between the $X$ and $Y$ basis for each sample. 

As ansatz for DualQITE, we use problem-inspired
circuits with pairwise CNOT couplings and $r=2$ repetitions of rotation and entanglement layers, plus a final rotation layer, see Appendix~\ref{app:su2_circuit} for a circuit diagram.
For evolutions of product states $\ket{\pm}$ in the $X$ basis, the rotation layers are single qubit $R_\mathrm{Y} R_\mathrm{Z}$ gates, and for the states
$\ket{\pm i}$ in the $Y$ basis, the layers implement $R_\mathrm{X} R_\mathrm{Z}$ gates.
The initial product states $\ket{\phi_m(0)}$ are prepared by setting the parameters in the final layer rotation layer of the ansatz as follows,
\begin{equation*}
    \begin{aligned}
        \ket{\pm} &\rightarrow R_\mathrm{Y}\left(\frac{\pm \pi}{2}\right) R_\mathrm{Z}(0), \\
        \ket{+i} & \rightarrow R_\mathrm{X}\left(\frac{\pi}{2}\right) R_\mathrm{Z}(\pi), \\
        \ket{-i} & \rightarrow R_\mathrm{X}\left(\frac{\pi}{2}\right) R_\mathrm{Z}(0).
    \end{aligned}
\end{equation*}
Each energy sample is evaluated with $1024$ measurements per basis.
The optimization problem in DualQITE is solved with a time perturbation $\delta\tau=0.01$ and gradient descent with a learning rate of $\eta=0.1$ and $100$ iterations in the first timestep, followed by $10$ iterations in the following, warmstarted timesteps.
We integrate with a fixed timestep of $\Delta_t = 0.01$.

\begin{figure}[htp]
    \centering
    \includegraphics[width=\linewidth]{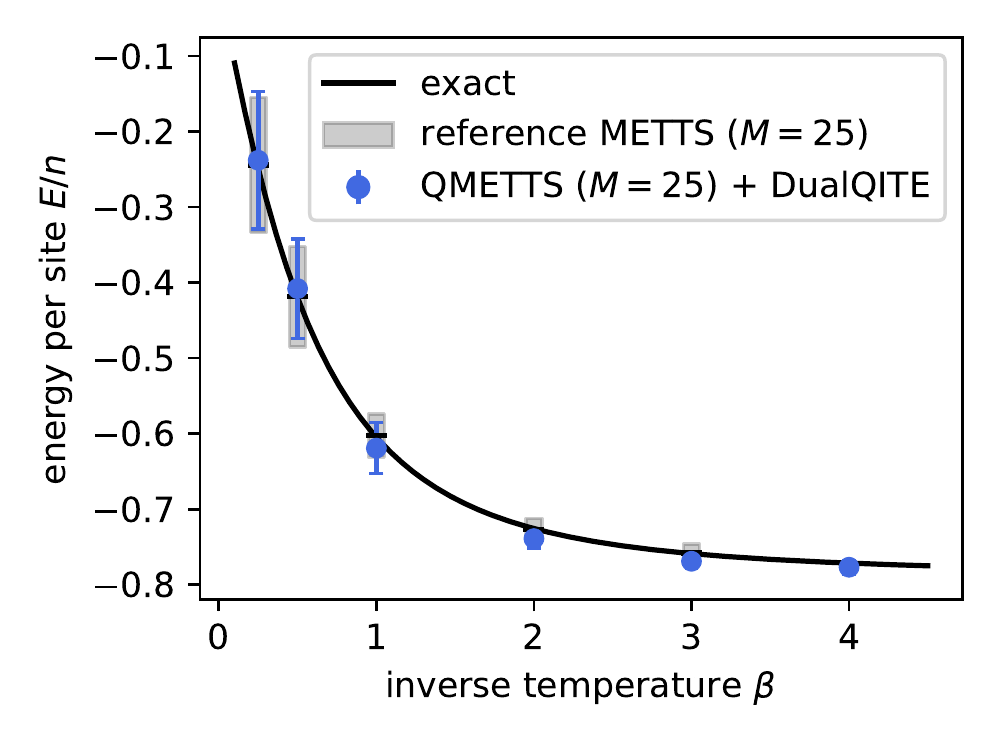}
    \caption{Energy per site for the Heisenberg model on a 6-spin chain, comparing mean and standard deviation of  QMETTS with DualQITE (blue circle and errorbars) with a reference METTS implementation (black line and grey shade).}
    \label{fig:energy_per_site}
\end{figure}

Figure~\ref{fig:energy_per_site} shows the estimated energy per site, along with the standard deviation of the samples, for different inverse temperatures $\beta$. For the alternating $X-Y$ basis, the Markov chain converges quickly and $M=25$ samples suffice for an accurate estimate of the observable.
For the imaginary-time evolution, we compare DualQITE with the same settings as in the previous sections to an exact evolution performed with matrix exponentials. It shows that using the dual method allows to reliably reproduce the mean and standard deviation of the Markov chain samples compared to the exact reference METTS.

\section{Real-time evolution}
\label{sec:real}

The focus of this paper is on imaginary-time evolution as, to date, no other QGT-free time evolution algorithms exist in this setting.
For real-time evolution, p-VQD has a similar structure as our algorithm and solves an optimization problem rather than evaluating the QGT.
However, there are key differences to the dual algorithm applied to real-time evolution.

The p-VQD algorithm \cite{barison_pvqd_2021} projects a single Suzuki-Trotter step onto the circuit model by solving the following optimization problem:
\begin{equation*}
    \vec\theta(t + \Delta_t) = \argmax{\vec\theta'} \big|\braket{\phi(\vec\theta') | e^{-iH\Delta_t} | \phi(\vec\theta(t))}\big|^2.
\end{equation*}
For Hamiltonians with many Pauli terms or long-range interactions, such as those arising in molecular dynamics, the single step might already lead to large circuits with non-local gates. 
While DualQRTE requires an LCU method to evaluate the imaginary part of the energy and state gradients, see Eq.~\eqref{eq:b_real}, this only adds a single entangling gate, compared to a full Suzuki-Trotter step is required.
Furthermore, our dual time evolution allows the evaluation of error bounds at almost no additional cost, which is not possible in p-VQD.
Due to these differences, this section presents DualQRTE: the dual time evolution for real-time evolution.

\subsection{Heisenberg model}

We present the real-time evolution under the Heisenberg Hamiltonian of Eq.~\eqref{eq:heisencomb} on a linear chain with $n=4$ spins with parameters $J=1/4,\; g=-1$. 
As variational model, we use a circuit with alternating Pauli-$X$ and Pauli-$Y$ rotation layers, and Pauli-$ZZ$ entangling gates that reflect the connectivity of the spins.
The circuit structure is visualized in Appendix~\ref{app:real_heisen} and, in this experiment, all algorithms use $r=3$ repetitions of the rotation as well as entangling gates. To prepare the initial state, $\ket{+}^{\otimes 4}$, we set the parameters of the final Pauli-$Y$ rotations to $\pi/2$ and the rest to 0.

During the evolution, we track the average magnetization in the $X$ and $Z$ direction,
\begin{equation*}
    \braket{X} = \frac{1}{n} \sum_{i=1}^n \braket{X_i}, ~~
    \braket{Z} = \frac{1}{n} \sum_{i=1}^n \braket{Z_i}.
\end{equation*} 
Since this Heisenberg Hamiltonian preserves the qubit excitations, and the initial state is the equal superposition, the $\braket{Z}$ expectation value should remain $0$ throughout the evolution. 

\begin{figure}
    \centering
    \includegraphics[width=\linewidth]{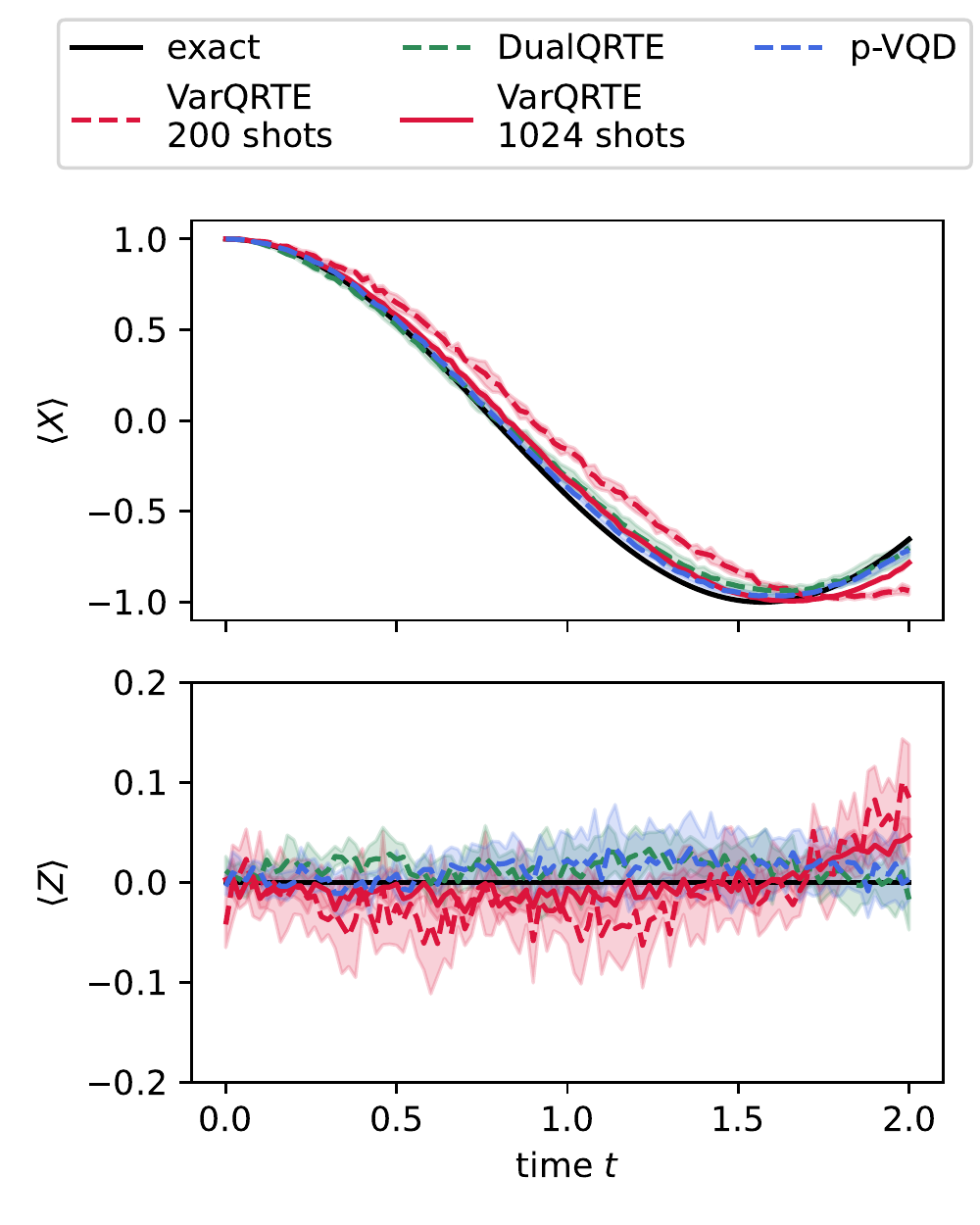}
    \caption{Average magnetization in $X$ and $Z$ direction as tracked by different variational algorithms.}
    \label{fig:magnetization}
\end{figure}

The results of the different time evolution algorithms for an integration time of $T=2$ and timestep $\Delta_t = T/100$ are presented in Fig.~\ref{fig:magnetization}.
Both DualQRTE and p-VQD accurately track the observables using only 200 shots per circuit. With the same resources, VarQRTE, on the other hand, has lower accuracy and we need to use 1024 shots per circuit to match the result of the optimization-based algorithms.

\subsection{Error bounds}

In variational real-time evolution, the model error in terms of Bures distance $D_B$ due to restriction to the variational manifold can be expressed as~\cite{zoufal_errorbounds_2021}
\begin{equation}
    \begin{aligned}
        \dot\varepsilon_M &:= \left\|\sum_{k=1}^d \dot\theta_k \ket{\partial_k \phi(\vec\theta)} + i H\ket{\phi(\vec\theta)}\right\|_2^2 \\
                          &= \mathrm{Var}(H | \phi(\vec\theta)) + \dot{\vec\theta}^T g(\vec\theta) \dot{\vec\theta} - 2 \dot{\vec\theta}^T \vec{b}(\vec\theta),
    \end{aligned}
\end{equation}
where we set $\hbar \equiv 1$.
Integrating this error rate provides an upper bound on the Bures distance, that is
\begin{equation*}
    D_B(\phi(\vec\theta(T)), \Psi(T)) \leq \int_0^T \dot\varepsilon_M(t) \mathrm{d} t, 
\end{equation*}
where $\ket{\Psi(t)}$ is the exact time-evolved state and the time-dependence of $\varepsilon_M$ is due to the time-dependence of the parameters $\vec\theta = \vec\theta(t)$.

Up to the variance $\mathrm{Var}(H | \phi(\vec\theta)) = \braket{\phi(\vec\theta)|H^2|\phi(\vec\theta)} - (\braket{\phi(\vec\theta)|H|\phi(\vec\theta)})^2$ , this error is proportional to the loss function used in DualQRTE. By using the same 
expansion $\dot{\vec\theta} = \vec{\delta\theta} / \delta\tau$ and using the infidelity to approximate the inner product with respect to the geometric tensor, we 
can rewrite the error as
\begin{equation*}
    \begin{aligned}
    \dot\varepsilon_M &= \mathrm{Var}(H | \phi(\vec\theta)) + \frac{1 - F(\vec\theta, \vec\theta + \vec{\delta\theta})}{(\delta\tau)^2} - \frac{2 \vec{\delta\theta}^T \vec{b}(\vec\theta)}{\delta\tau} + \mathcal{O}(\delta\tau) \\
    &= \mathrm{Var}(H | \phi(\vec\theta)) + \frac{2 \mathcal{L}(\vec{\delta\theta})}{(\delta\tau)^2} + \mathcal{O}(\delta\tau).
    \end{aligned}
\end{equation*}
Note that the error scales linearly in time perturbation $\delta\tau$ as the infidelity approximation has a cubic error term \cite{stokes_qng_2020}, which is divided by the square of the perturbation. If we, for example, use a forward Euler rule with timestep $\Delta_t$ to integrate the variational error, the integration error scales as $\mathcal{O}(\Delta_t + T\delta\tau)$. This highlights the importance of differentiating between the timestep $\Delta_t$ for the integration, and the time-perturbation $\delta\tau$ to approximate the derivative.

\begin{figure}[htp]
    \centering
    \includegraphics[width=\linewidth]{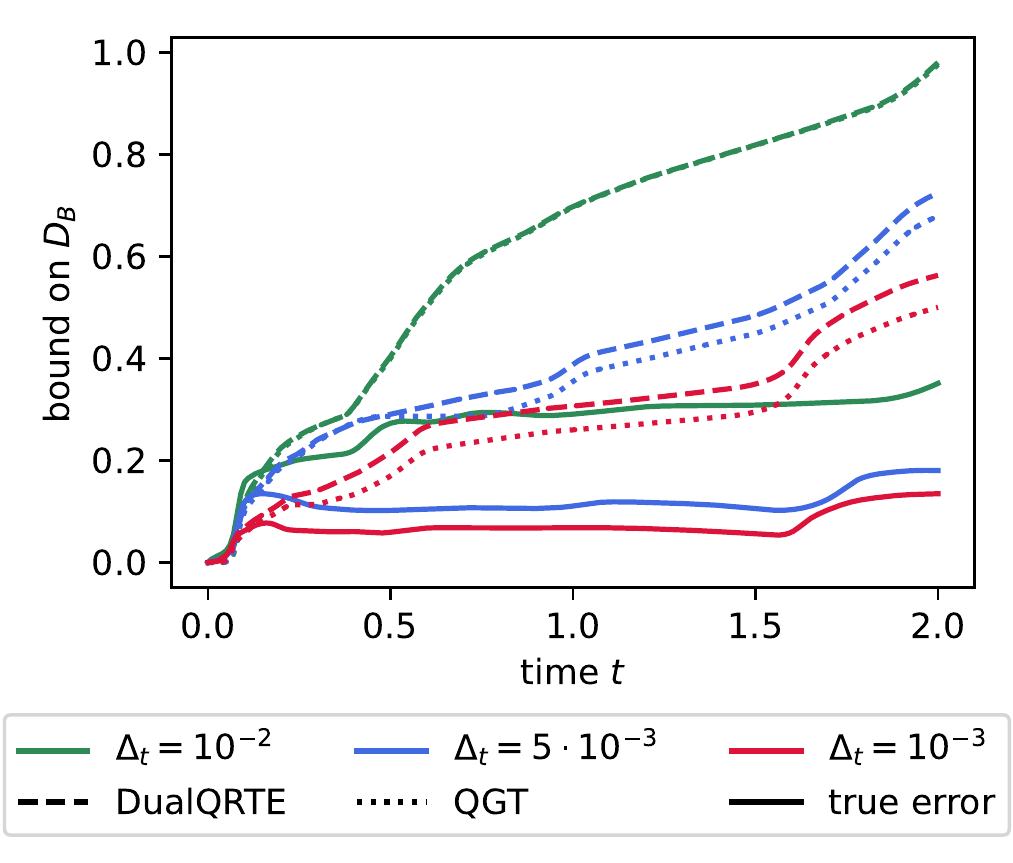}
    \caption{Error of the real-time evolution in Bures distance, plus error bounds obtained with VarQRTE and DualQRTE.}
    \label{fig:errorbounds}
\end{figure}

In Fig.~\ref{fig:errorbounds}, we show the error bounds along with the true error for the time evolution of the Heisenberg model. The bounds are computed for different timesteps $\Delta_t$ for VarQRTE, and for DualQRTE for a fixed time perturbations $\delta\tau = 10^{-3}$ in exact simulations.
Firstly, we can verify that the error bounds hold.
Secondly, the larger the timestep relative to the time-perturbation, the more accurate the approximation of the dual time evolution, as the error $\mathcal{O}(\Delta_t + T\delta\tau)$ is dominated by the integration error. 

\section{Conclusion}
\label{sec:conclusion}

In this paper, we present a novel algorithm for variational quantum time evolution that does not require the evaluation of the QGT, but instead solves a dual optimization problem in each timestep. The proposed dual time evolution algorithm, DualQTE, is particularly interesting for imaginary-time evolution, as there is currently no alternative variational algorithm able to circumvent the $\mathcal{O}(d^2)$ cost of VarQITE. For real-time evolution, p-VQD also offers an optimization-based approach by projecting a single Trotter step onto the variational form. In comparison, the dual time evolution has the advantage that no Suzuki-Trotter step has to be implemented, which could require deep circuits or non-local operations, depending on the Hamiltonian. Furthermore, our algorithm allows to evaluate variational error bounds \cite{zoufal_errorbounds_2021}, although how accurately they can be evaluated in the presence of shot noise remains an open question.

We demonstrated DualQTE for the imaginary-time evolution of a Heisenberg Hamiltonian on 12 qubits, 
and found that, in this setting, it requires about one order of magnitude less measurements to achieve the same accuracy as VarQITE. 
As a practical application of imaginary-time evolution, we calculated thermodynamic observables with the QMETTS algorithm and showed that the DualQITE is suitable to reproduce the sampling distributions.
Finally, we applied our algorithm to an illustrative example for real-time evolution, where it produced comparable results to p-VQD for the same amount of resources, while both algorithms outperformed VarQRTE.

In the presented experiments, we used standard gradient descent algorithms with a fixed learning rate.
We expect that the performance could be further improved by using more advanced optimization schemes, or methods that also take into account information from previous iterations.
Another possible improvement would be a suitable termination criterion for noisy evaluations of the loss function.
As for other optimization-based time evolution algorithms, such as p-VQD, 
it remains challenging to accurately measure the fidelity in the presence of hardware noise.

In conclusion, the proposed DualQTE is an efficient variational algorithm for quantum time evolution that does not suffer from the quadratic complexity of evaluating the QGT.
This cost reduction enables scaling imaginary-time evolution to larger, practically relevant system sizes and allows the simulation and demonstration of a wide variety of important tasks such as Gibbs state preparation, mixed time evolution, or the evaluation of thermodynamic observables. 
Improving the resource requirements for near-term algorithms is an important step for scaling demonstrations to the full size of today's quantum computers and work towards practical applications.

\section{Acknowledgements}

We thank Christa Zoufal, Stefano Barison, Almudena Carrera Vazquez, David Sutter, Caroline Tornow, Laurin Fischer and Daniel Egger for insightful conversations on this project.

We acknowledge the use of IBM Quantum services for this work. The views expressed are those of the authors, and do not reflect the official policy or position of IBM or the IBM Quantum team.

IBM, the IBM logo, and ibm.com are trademarks of International Business Machines Corp., registered in many jurisdictions worldwide. Other product and service names might be trade- marks of IBM or other companies. The current list of IBM trademarks is available at \url{https://www.ibm.com/legal/copytrade}.

\bibliography{refs} 

\appendix
\onecolumngrid

\section{Runtime estimates of variational time evolution}\label{app:runtime_estimates}

The benchmark in Sec.~\ref{sec:complexity_benchmark} provides a scaling for the total number of measurements $N$ required by VarQITE and DualQITE, which allows a runtime estimation on the algorithms on quantum hardware.
In this estimation we neglect the overhead of classical processors and assume a superconducting quantum computer with a basis gate set including $\sqrt{X}$, $R_\mathrm{Z}$ and CX gates, as reported by several IBM Quantum backends, for example. This gate set allows to compile any sequence of single qubit gates into two $\sqrt{X}$ gates and three virtual $R_\mathrm{Z}$ gates.
For an $n$-qubit simulation of the Heisenberg model and the considered circuit model (see Fig.~\ref{fig:efficient_su2}) with $r$ repetitions, the time for a single measurement can then be approximated as 
\begin{equation}
    t_\text{shot} = 2r t_\text{CX} + 2(r + 1) t_{\sqrt{X}} + t_\text{meas} + t_\text{reset},
\end{equation}
where $t_\text{CX}$ is the duration of a CX gate, $t_{\sqrt{X}}$ the duration of a $\sqrt{X}$ gate, $t_\text{meas}$ the time of a measurement and $t_\text{reset}$ the time to reset the qubits for the next execution. Since the $R_\mathrm{Z}$ gates are virtual they do not contribute to the runtime.
The total runtime is then estimated by $N t_\text{shot}$.

Depending on the architecture and the gate decomposition the duration and fidelity of single- and two-qubit operation, as well as measurements, varies on superconducting qubit chips~\cite{kjaergaard_superconducting_2020}. 
Here we use gate times of $t_\text{CX} = 451ns$, $t_{\sqrt{X}} = 36ns$ and $t_\text{meas} = 860ns$ as reported by \texttt{ibm\_peekskill} (v2.6.5), which is an IBM Quantum Falcon processors~\cite{ibm_quantum}.
For shallow circuits in particular, the time to reset qubits for the following execution is a crucial bottleneck. 
The reset operation can, for example, be implemented by waiting 5-10 T1 times and let the qubits decay to the computational ground-state, but with T1 times of the order of $400\mu s$ the reset via relaxation is orders of magnitude slower than the other circuit operations.
Active resets instead measure the qubit state and apply an $X$-operation conditionally if the state $\ket{1}$ is measured.
This technique allows to reduce the reset times to typically $50$ to $250\mu s$ on IBM hardware~\cite{tornow_restless_2022}, which, however, still dominates the overall runtime for the considered circuits.
By using a second excited state it is possible to implement reset schemes with $500ns$ to $2\mu s$~\cite{egger_reset_2018, magnard_reset_2018} and we therefore use $t_\text{reset} = 2\mu s$ in our estimation.

\section{Derivation via quantum natural gradient descent}\label{app:qng_derivation}

VarQITE is inherently connected to the quantum natural gradient (QNG) algorithm~\cite{stokes_qng_2020}. In fact, this connection is a motivation for the convergence of the QNG as imaginary-time evolution is guaranteed to converge to the ground state, if there is sufficient initial overlap with it.

With a forward Euler integration the VarQITE update rule is
\begin{equation*}
    \vec\theta^{(t+1)} = \vec\theta^{(t)} + \Delta_t g^{-1}(\vec\theta^{(t)})\left(-\frac{\vec\nabla E(\vec\theta^{(t)})}{2}\right).
\end{equation*}
This coincides with the QNG update step for the loss function $\ell(\vec\theta) = E(\vec\theta)/2$ and a learning rate of $\eta = \Delta_t$,
\begin{equation}\label{eq:qng}
    \vec\theta^{(t+1)} = \vec\theta^{(t)} - \eta g^{-1}(\vec\theta^{(t)}) \vec\nabla\ell(\vec\theta^{(t)}).
\end{equation}
The natural gradients step can be expressed in a dual formulation as 
\begin{equation*}
    \vec\theta^{(t+1)} = \argmin{\vec\theta} \braket{\vec\nabla\ell(\vec\theta^{(t)}, \vec\theta - \vec\theta^{(t)}} + \frac{1}{2 \eta} d^2(\vec\theta, \vec\theta^{(t)}),
\end{equation*}
with a distance metric $d$. In this equation we see that the update step is going into the opposite direction of the gradient $\vec\nabla\ell$, while the magnitude is limited by the distance metric and the learning rate.

Standard gradient descent uses the model-agnostic $\ell_2$ norm as distance metric. Natural gradients on the other hand limit the update step by the amount of change it induces in the model. To measure the induced change the metric $d$ is chosen to be the Fubini-Study metric, which, as shown in Ref.~\cite{stokes_qng_2020}, if locally approximated, yields the QGT:
\begin{equation*}
    \begin{aligned}
    d^2(\phi(\vec\theta), \phi(\vec\theta + \vec{\delta\theta})) &= \arccos^2|\braket{\phi(\vec\theta)|\phi(\vec\theta + \vec{\delta\theta})}| \\
    &= 1 - |\braket{\phi(\vec\theta)|\phi(\vec\theta + \vec{\delta\theta})}|^2 + \mathcal{O}(||\vec{\delta\theta}||_2^4) \\
    &= \braket{\vec{\delta\theta}, g(\vec\theta)\vec{\delta\theta})} + \mathcal{O}(||\vec{\delta\theta}||_2^3).
    \end{aligned}
\end{equation*}
The formulation in Eq.~\eqref{eq:qng} is then obtained by solving the minimization problem.

To circumvent the explicit evaluation of the QGT the natural gradient update can instead be calculated without the quadratic local approximation, and instead solve the optimization problem directly. If we use the infidelity as distance metric and replace the loss function gradient by the evolution gradient $\vec\nabla\ell(\vec\theta) = \vec\nabla E(\vec\theta)/2 = -\vec{b}(\vec\theta)$, we obtain the same update rule as the main text
\begin{equation*}
    \vec\theta^{(t+1)} = \argmin{\vec\theta} -\braket{\vec{b}(\vec\theta^{(t)}), \vec\theta - \vec\theta^{(t)}} + \frac{1 - F(\vec\theta, \vec\theta^{(t)})}{2 \Delta_t}.
\end{equation*}

\section{Illustrative example}\label{app:illustrative_example}

\begin{figure*}[htp]
    \centering
    \includegraphics[width=0.4\textwidth]{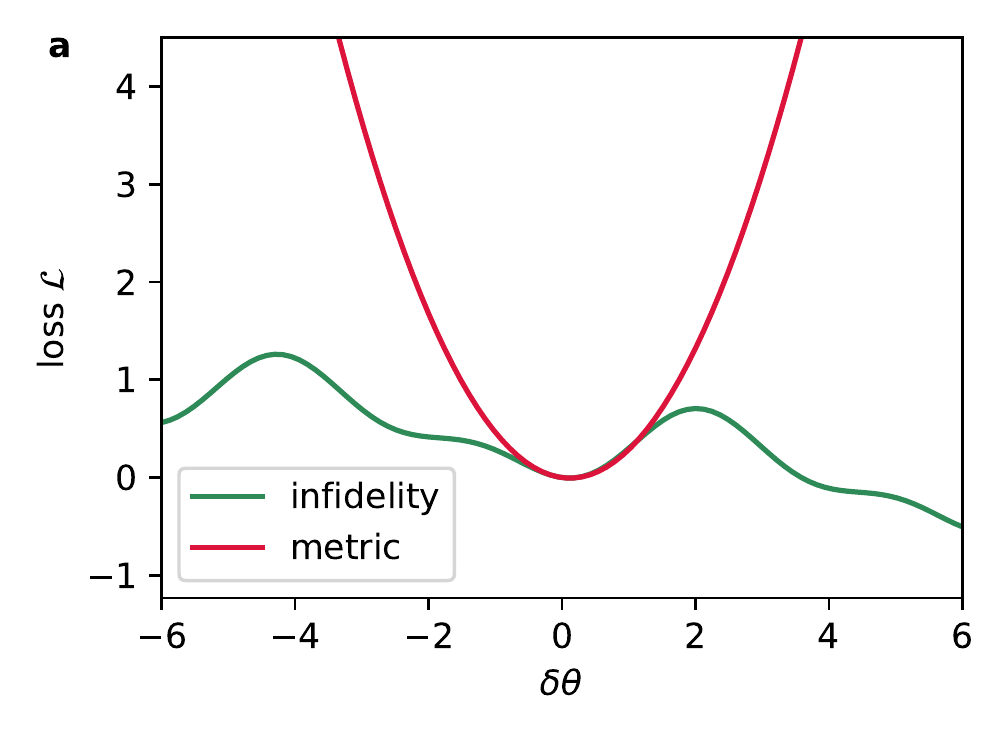}
    \includegraphics[width=0.4\textwidth]{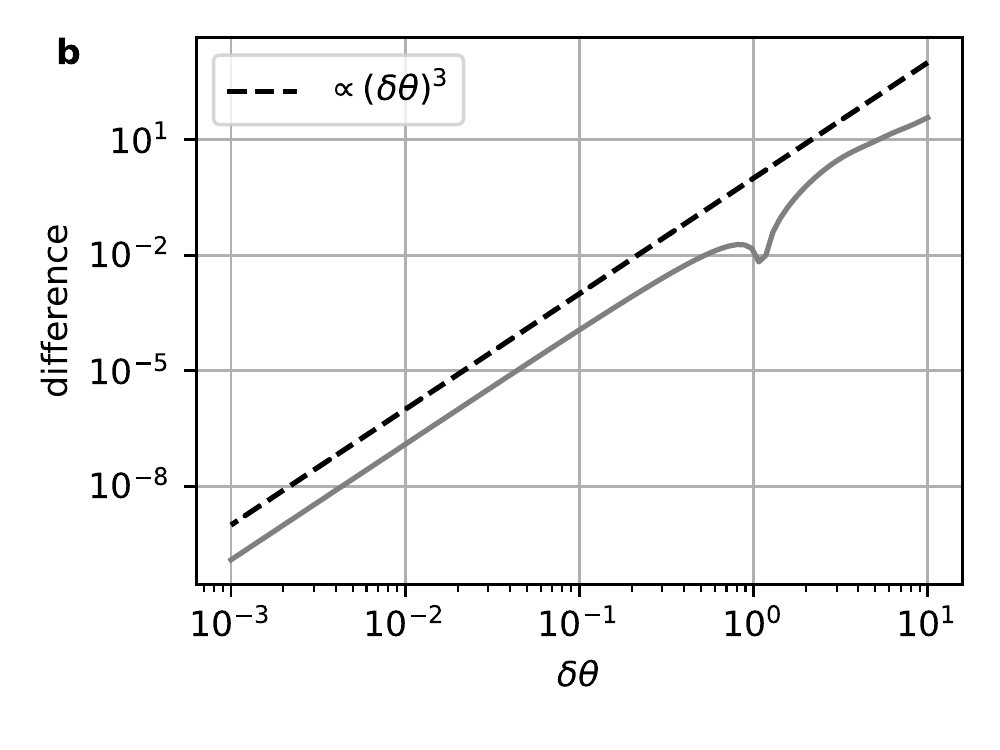}
    \caption{(a) Values of the loss function $\mathcal{L}$ for evaluation with the QGT metric, and with introduced infidelity approximation.
    (b) Difference of the QGT metric and infidelity as function of the perturbation $\delta\theta$.}
    \label{fig:illustrative}
\end{figure*}

For an intuitive understanding of the approximations of the QGT norm,
we investigate an illustrative example with the variational model $\ket{\phi(\theta)} = R_\mathrm{Z}(\theta) R_\mathrm{Y}(\theta)\ket{0}$, the Hamiltonian $H = Z$ and a timestep of $\delta\tau = 1/2$. In Fig.~\ref{fig:illustrative}(a) we compare the exact values of the loss function $\mathcal{L}$ for imaginary-time evolution around $\theta=\pi/4$ obtained by using the metric $\braket{\delta\theta, g(\theta)\delta\theta}$ or the infidelity $1 - F(\theta, \theta + \delta\theta)$ as norm.

At $\delta\theta = 0$ the approximation is exact and in the vicinity the difference scales as $(\delta\theta)^3$, see also Fig.~\ref{fig:illustrative}(b). 
Note, that the infidelity is periodic and bounded in $[0, 1]$ but the linear term $b^T\delta\theta$ is unbounded, which leads to the fact that the minimum of the infidelity-based loss function close to $\delta\theta = 0$ is not the global minimum.
This is well visible in Fig.~\ref{fig:illustrative}(a) where the infidelity-based loss function achieves lower values for large $\delta\theta$ than the minimum of the QGT close to $\delta\theta = 0$.
Since we aim to find the same minimum as the QGT-based loss function using a local optimization routine, such as gradient descent, is crucial for the dual time evolution.

\subsection*{Impact of the time perturbation}

The approximation error scales with the norm of $\vec{\delta\theta}$ and, therefore, solving for the update step $\dot{\vec\theta} = \vec{\delta\theta} / \delta\tau$ with a smaller time perturbation $\delta\tau$ should result in a smaller error in the update step. Remembering the definition of the loss function
\begin{equation*}
    \mathcal{L}(\vec{\delta\theta}) = \frac{1 - F(\vec\theta, \vec\theta+\vec{\delta\theta})}{2} - \delta\tau \cdot \vec{b}^T(\vec\theta) \vec{\delta\theta},
\end{equation*}
we see that a smaller $\delta\tau$ moves the minimum closer to the minimum of the infidelity at $\vec{\delta\theta} = \vec{0}$, leading to a smaller approximation error. 
Since the fidelity is bounded but the linear part $\vec{b}^T(\vec\theta) \vec{\delta\theta}$ is not, there is a maximum feasible range for the value of $\delta\tau$. 
A necessary condition for the existence of the minimum is that the gradient of the loss function vanishes, $\vec\nabla\mathcal{L} = 0$, which requires
\begin{equation}
    \forall i \in \{1, \dots, d\}: \frac{1}{2}\frac{\partial}{\partial(\delta\theta)_i} F(\vec\theta, \vec\theta + \vec{\delta\theta}) = -\delta\tau \cdot b_i(\vec\theta).
\end{equation}
For a circuit with unique parameters and only Pauli rotations gates, the gradient of the fidelity can be bounded via the parameter-shift rule to be in $[-1/2, 1/2]$ (see also Appendix~\ref{app:sampling_error}). Thus, a necessary condition for the timestep perturbation is
\begin{equation}
    \forall i \in \{1, \dots, d\}: \delta\tau \in \left[\frac{-1}{4|b_i(\vec\theta)|}, \frac{1}{4|b_i(\vec\theta)|}\right],
\end{equation}
which can be generalized to circuits with repeated parameters or other than Pauli gates. 
Note that this is only a necessary and not a sufficient condition for the existence of a minimum since, depending on the circuit structure, the fidelity gradient may not support the full range $[-1/2, 1/2]$.

\begin{figure*}[htp]
    \centering
    \includegraphics[width=\textwidth]{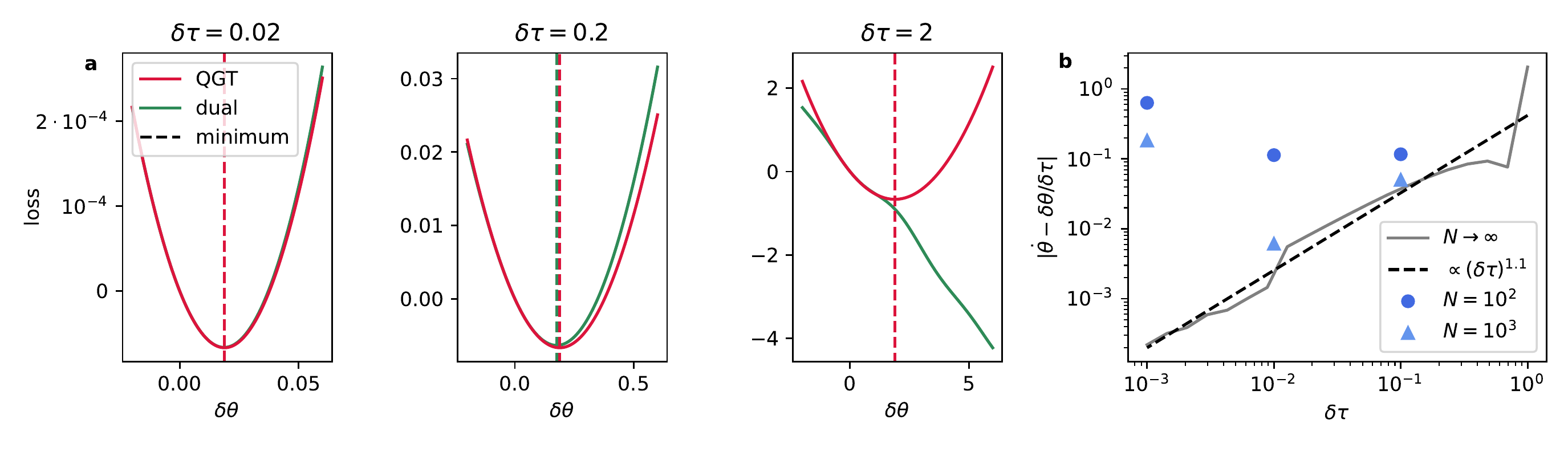}
    \caption{(a) Loss landscapes and optimal solutions of the original, QGT-based loss function and the dual loss function for different $\delta\tau$. (b) Error in calculating the parameter derivative $\dot\theta$ depending on $\delta\tau$ and the number of measurements $N$.}
    \label{fig:dtau_scaling}
\end{figure*}

In Fig.~\ref{fig:dtau_scaling}(a) we visualize the impact of $\delta\tau$ on the loss landscape. For small time perturbations the QGT-based and dual loss landscapes almost coincide, but if $\delta\tau$ is chosen too large the dual loss function has no minimum.
If the loss function can be evaluated exactly, choosing $\delta\tau$ as small as possible therefore minimizes the approximation error. In Fig.~\ref{fig:dtau_scaling}(b), we find the the error in the parameter derivative $\dot{\vec\theta} = \vec{\delta\theta} / \delta\tau$ scales approximately as $\mathcal{O}(\delta\tau)$.
In practice, however, the loss function is subject to measurement noise and errors in the solution $\vec{\delta\theta}$ are amplified by $1/\delta\tau$. Hence, for a finite number of measurements there is a trade-off between QGT approximation error and controlling the noise amplification.

\section{Bound the sample complexity of VarQTE}\label{app:sampling_error}

In this section, we present the derivation on the upper bound of the sample complexity of VarQTE and DualQTE.  
The target error is measured in integrated Bures distance,
\begin{equation}
    \varepsilon_S = \frac{1}{T}\int_0^T \sqrt{2(1 - |\braket{\phi(\vec\theta) | \phi(\tilde{\vec\theta})}|}\mathrm{d} t.
\end{equation}
Assuming a forward Euler integration, the Bures distance can be formulated in terms of the QGT as
\begin{equation}\label{eq:bures_as_l2}
    \begin{aligned}
        \varepsilon_S &= \frac{1}{T}\int_0^T \sqrt{2 \left(1 - \sqrt{1 - \Delta_t^2 \Delta\dot{\vec\theta}^T g(\vec\theta) \Delta\dot{\vec\theta} }\right)} \mathrm{d} t\\
        &= \frac{1}{T}\int_0^T \Delta_t \sqrt{\Delta\dot{\vec\theta} g(\vec\theta)\Delta\dot{\vec\theta}} \mathrm{d}t\\
        &\leq \frac{1}{T} \int_0^T \Delta_t \|g(\vec\theta)\|_2 \|\Delta\dot{\vec\theta}\|_2 \mathrm{d}t \\
        &\leq \Delta_t \sqrt{\lambda_\text{max}} \|\Delta\dot{\vec\theta}_\mathrm{max}\|_2,
    \end{aligned}
\end{equation}
where we introduced $\Delta\dot{\vec\theta} = \tilde{\dot{\vec\theta}} - \dot{\vec\theta}$, $\lambda_\text{max} \geq 0$ is a bound on the largest eigenvalue of $g$ for any parameter value $\vec\theta$ and, similarly, $\|\Delta\dot{\vec\theta}_\mathrm{max}\|_2$ an upper bound on the norm $\Delta\dot{\vec\theta}$.
In the first line, we dropped $\mathcal{O}(\Delta_t^3)$ error terms, in the second line we used a first order Taylor-expansion and in the last line we use the definition of the operator norm to bound the inner product of $\vec{\delta\theta}$ in the metric of $g$.

\subsection{VarQTE}

In each VarQTE step we solve a linear system for the update step, where the measurements of the QGT and evolution gradient are subject to sampling error. We define the noisy quantities as $\tilde g(\vec\theta) = g(\vec\theta) + \Delta g(\vec\theta)$ and $\tilde{\vec b}(\vec\theta) = \vec{b}(\vec\theta) + \Delta \vec{b}(\vec\theta)$ and, then, solve the noisy linear system
\begin{equation}
    \tilde g(\vec\theta) \tilde{\dot{\vec\theta}} = \tilde{\vec b}(\vec\theta),
\end{equation}
with the noisy update $\tilde{\dot{\vec\theta}} = \dot{\vec\theta} + \Delta\dot{\vec\theta}$.
To stabilize the linear system and ensure the QGT and it's estimate are invertible, we assume a regularization of $g$ and $\tilde g$ in form of a diagonal shift $\delta_c$.
This shift is a trade-off of stability and bias, which is also discussed in Ref.~\cite{gacon_qnspsa_2021}, Appendix D.

We can write the error in the update step using the difference of the noisy and exact linear system solutions, as
\begin{equation}\label{eq:deltatheta_bound}
    \begin{aligned}
        \|\Delta\dot{\vec\theta}\|_2 &= \|(g + \Delta g)^{-1}(\vec b + \Delta \vec b) - g^{-1}\vec b\|_2  \\
        &\approx \|(g^{-1} - g^{-1}\Delta g g^{-1})(\vec b + \Delta\vec b) - g^{-1} \vec b \|_2 \\
        &= \|g^{-1}\Delta \vec b - g^{-1} \Delta g g^{-1} \vec b - g^{-1} \Delta g^{-1} g^{-1}\Delta \vec b \|_2  \\
        &\approx \|g^{-1}\Delta \vec b - g^{-1}\Delta g \dot{\vec\theta} \|_2 \\
        &\leq \|g^{-1}\|_2 \left(\|\Delta\vec b\|_2 + \|\Delta g\|_2 \|\dot{\vec\theta}\|_2\right),
    \end{aligned}
\end{equation}
where we dropped the explicit parameter dependence for legibility.
In the second line we used the Neumann series to approximate $(g + \Delta g)^{-1} = g^{-1} - g^{-1}\Delta g g^{-1} + \mathcal{O}(\|\Delta g\|_2^2 \|g^{-1}\|_2^3)$ and dropped quadratic error terms on the fourth line.
In the following we derive upper bounds on the maximal value of the individual contributions in the error bound, such that we finally obtain a bound $\|\Delta\dot{\vec\theta}_\mathrm{max}\|$ on the error in the update step.

\paragraph{Spectrum of $g$}

Each QGT entry can be computed as \cite{gacon_qnspsa_2021}
\begin{equation}
    g_{ij}(\vec\theta) = -\frac{1}{2} \partial_i\partial_j F(\vec\theta', \vec\theta) \Bigg\vert_{\vec\theta' = \vec\theta}.
\end{equation}
For a circuit with unique, non-interacting parameter and only plain Pauli rotation gates $R_P(\theta)$, we can use the parameter-shift rule \cite{schuld_gradients_2019}
to write the entry as
\begin{equation}
    g_{ij}(\vec\theta) = -\frac{1}{2} \frac{F^{(++)}_{ij} - F^{(+-)}_{ij} - F^{(-+)}_{ij} + F^{(--)}_{ij}}{4},
\end{equation}
where we abbreviated $F^{(\pm\pm)}_{ij} = F(\vec\theta, \vec\theta \pm \vec{e}_i \pi/2 \pm \vec{e}_j \pi/2)$ for the $i$th unit vector $e_i$ (and $j$th unit vector $e_j$).
Since the fidelity is in $[0, 1]$ we can bound each entry by
\begin{equation}
    -\frac{1}{4} \leq g_{ij}(\vec\theta) \leq \frac{1}{4},
\end{equation}
for any value of $\vec\theta$.
Gershgorin's circle theorem tells us that the maximal eigenvalue of $g$ is bounded from above by the maximal sum over the columns or rows, which in this case is achieved by setting all elements of a column/row to $1/4$. This gives the bound
\begin{equation}
    \lambda_\text{max} \leq \sum_{i=1}^{d} \frac{1}{4} = \frac{d}{4}.
\end{equation}

This bound can be generalized to circuits with coefficients or repeated parameters by applying the chain and product rules. 
For example, for a coefficient-free circuit where parameters can be repeated up to $m$ times, the bound becomes $md/4$.

\paragraph{Norm of the update step}

The update step can be bounded as $\|\dot{\vec\theta}\|_2 \leq \|g^{-1}\|_2 \|b\|_2$, where $\|g^{-1}\|_2 \leq \delta_c^{-1}$. The evolution gradient can 
be bounded using the parameter-shift rule, under the circuit structure assumptions as the previous section.
Each element in the gradient is bounded by
\begin{equation}
    \begin{aligned}
        |b_i| = \frac{|E^{(+)}_i - E^{(-)}_i|}{2} \leq \frac{|E^{(+)}_i| + |E^{(-)}_i|}{2} \leq E_\text{max},
    \end{aligned}
\end{equation}
where $E^{(\pm)}_i = E(\vec\theta \pm \vec{e}_i \pi/2)$ and $E_\text{max}$ is the absolute maximum system energy.
The norm over all elements is then $\|b\|_2 \leq \sqrt{d} E_\text{max}$, leading to an overall bound of
\begin{equation}\label{eq:dottheta_bound}
    \|\dot{\vec\theta}\|_2 \leq \frac{\sqrt{d} E_\text{max}}{\delta_c},
\end{equation}
for any parameter value $\vec\theta$.

\paragraph{Sampling errors}

Since the measurement noise is unbiased, the random variable $\Delta g = \tilde g - g$ has zero mean with i.i.d. entries. This allows to apply Latala's theorem \cite{latala_bound_2005}, which states that 
\begin{equation}
    \mathbb{E}[\|\Delta g\|_2] \leq C \left(\max_{i} \sqrt{\sum_{j=1}^d \mathbb{E}[(\Delta g)_{ij}^2]} + \max_{j} \sqrt{\sum_{i=1}^d \mathbb{E}[(\Delta g)_{ij}^2]} + \sqrt[4]{\sum_{i, j=1}^d \mathbb{E}[(\Delta g)_{ij}^4]} \right),
\end{equation}
for some constant $C \in \mathbb{R}$.

Ref.~\cite{gentinetta_pegasos_2022} is concerned with the similar case of sampling the matrix $[F(x_i, x_j)]_{i,j=1}^d$ for a set of parameters $\{x_i\}_{i=1}^d$.
There, the matrix entries are Bernoulli distributed with probability $F(x_i, x_j)$. Using QGT representation as Hessian and applying the parameter-shift rule, we can see that the entries of $g_{ij}$ are Poisson binomial distributed~\cite{wang_number_1993} with probabilities $[F^{(++)}_{ij}, 1- F^{(+-)}_{ij}, 1-F^{(-+)}_{ij},F^{(--)}_{ij}]$ over a shifted support $[0, 1, 2, 3, 4] \rightarrow [-2, -1, 0, 1, 2]$.
Since the of this distribution are independent of the number of circuit parameters, it can be shown analogous to Ref.~\cite{gentinetta_pegasos_2022} that 
\begin{equation}
    \mathbb{E}[|(\Delta g)_{ij}|^2] = \mathcal{O}\left(\frac{1}{N}\right) \text{ and }
    \mathbb{E}[|(\Delta g)_{ij}|^4] = \mathcal{O}\left(\frac{1}{N^2}\right),
\end{equation}
which leads to a total bound of 
\begin{equation}
    \mathbb{E}[\|\Delta g\|_2] = \mathcal{O}\left(\sqrt{\frac{d}{N}}\right).
\end{equation}

The bound on $\|\Delta \vec b\|_2$ does not need to be tighter than $\|\Delta g\|_2 \|\dot{\vec\theta}\|_2$, which is straightforward to achieve via the sampling error.
Using the product rule we have
\begin{equation}
    \begin{aligned}
        |\Delta b_i| &= |\tilde b_i - b_i| = \frac{|\tilde E^{(+)}_i - \tilde E^{(-)}_i - E^{(+)}_i + E^{(-)}_i|}{2}  \\
        &\leq \frac{|\tilde E^{(+)}_i - E^{(+)}_i| + |\tilde E^{(-)}_i - E^{(-)}_i|}{2}  \\
        &= \mathcal{O}\left(\frac{\sqrt{\mathrm{Var}(E^{(+)}_i)} + \sqrt{\mathrm{Var}(E^{(-)}_i)}}{2\sqrt{N}}\right).
    \end{aligned}
\end{equation}
The variance of any state $\ket{\psi}$ can be upper bounded by
\begin{equation}
    \mathrm{Var}(E) = \braket{\psi|H^2|\psi} - E^2 \leq \braket{\psi|H^2|\psi} \leq E_\text{max}^2.
\end{equation}
Summing over all gradient elements we obtain
\begin{equation}
    \|\Delta\vec b_\mathrm{max}\|_2 = \mathcal{O}\left(\frac{\sqrt{d}E_\text{max}}{\sqrt{N}}\right).
\end{equation}

\paragraph{Final bound} 

Plugging the bounds in the previous paragraphs into Eq.~\eqref{eq:deltatheta_bound} and then into Eq.~\eqref{eq:bures_as_l2}, we obtain the final bound of
\begin{equation}
    \varepsilon_S \leq \mathcal{O}\left(\frac{d^{3/2} E_\text{max} \Delta_t}{\delta_c^2 \sqrt{N}}\right).
\end{equation}
The same asymptotic bound can be derived by performing a moment expansion on the expectation $\mathbb{E}[\dot{\vec\theta} - \tilde{\dot{\vec\theta}}]$.

As an example we investigate a simple product-state model, which allows to show the tightness of several of the above bounds.
We look a the first timestep of the $n$-qubit Hamiltonian $H = \sum_{i=1}^n Z_i$ with an ansatz that consists of a single layer of Pauli-$Y$ rotations, each with an individual parameter.
The initial state is $\ket{+}^{\otimes n}$ which is prepared by setting each of the parameters to $\pi/2$. Each expectation value is computed with $N=1000$ measurements 
and we use a regularization of $\delta_c = 10^{-2}$.
We then vary the number of qubits from $n=2$ to $10$ and measure the error term contributions over 10 averages, since $\Delta g$ and $\Delta\vec b$ are random variables. 

\begin{figure*}[htp]
    \centering
    \includegraphics[width=0.9\textwidth]{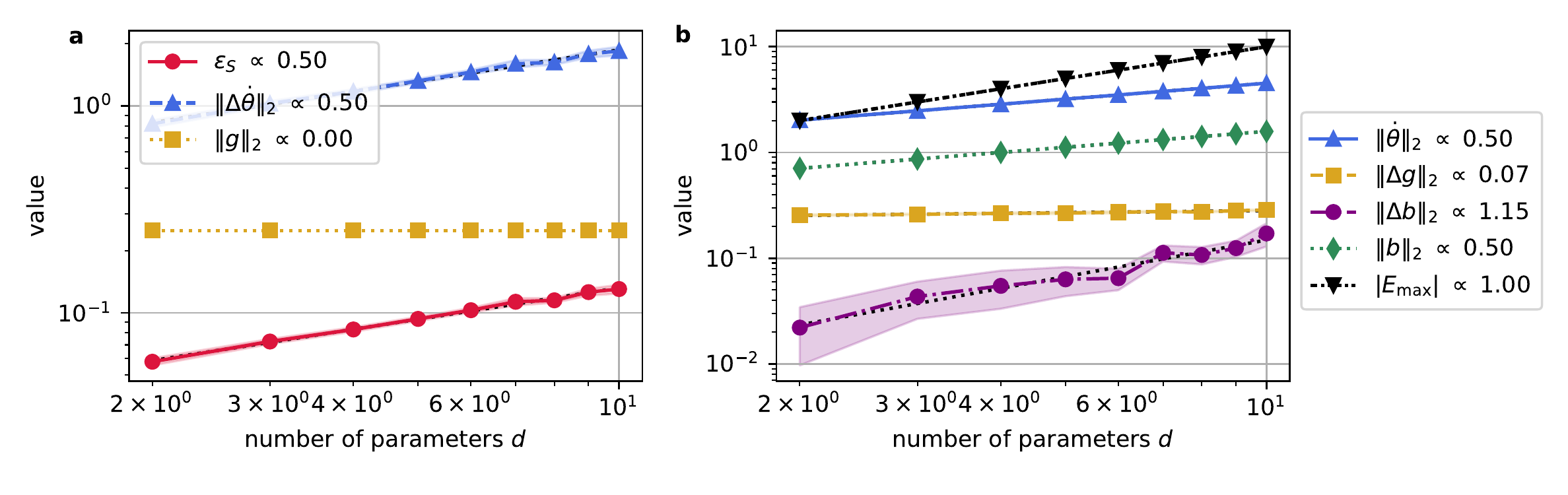}
    \caption{Scaling of different error bound contributions for a product state setting. The labels include the scaling with number of parameters, i.e. $\propto \alpha$ indicates a scaling with $d^\alpha$, where $d$ is the number of parameters.}
    \label{fig:error_scalings}
\end{figure*}

The QGT is measure of the correlation between the parameter derivatives and as there is not light-cone connecting any two parameterized gates in the product state ansatz, the QGT is a diagonal matrix. Its norm is therefore $\|g\|_2 = 1/4$ for any system size. With this restriction, we observe in Fig.~\ref{fig:error_scalings}(a) that the bound on the Bures metric in Eq.~\eqref{eq:bures_as_l2} is tight as $\varepsilon_S \propto \|\Delta\dot{\vec\theta}\|_2$.
While all bounds are obeyed, we observe that in particular the bound on $\|\dot{\vec\theta}\|_2$ is loose, since the bound in Eq.~\eqref{eq:dottheta_bound} scales with $\sqrt{d}E_\text{max} \propto d^{1.5}$, but we only observe a $d^{0.5}$ scaling. This bound could potentially be further improved by taking into account that the magnitude of the update step is bounded by the change induced of the evolution operator $\exp(-\Delta_t H)$, which is independent of the number of parameters $d$.

\subsection{DualQTE}

Assume we require $K$ steps to converge. Then the error in the update step $\vec{\delta\theta}$ is 
\begin{equation}
    \begin{aligned}
        \|\Delta(\vec{\delta\theta})\|_2 = \|\Delta(\vec{\delta\theta}^{(K)})\|_2 &= \|\widetilde{\vec{\delta\theta}}^{(K - 1)} - \eta\widetilde{\vec\nabla\mathcal{L}}(\widetilde{\vec{\delta\theta}}^{(K-1)}) - \vec{\delta\theta}^{(K-1)} + \eta\vec\nabla\mathcal{L}(\vec{\delta\theta}^{(K-1)}) \|_2\\
        &\leq \|\Delta(\vec{\delta\theta}^{(K-1)})\|_2 + \eta \|\widetilde{\vec\nabla\mathcal{L}}(\widetilde{\vec{\delta\theta}}^{(K-1)}) - \vec\nabla\mathcal{L}(\vec{\delta\theta}^{(K-1)}) \|_2 \\
        &\leq \|\Delta(\vec{\delta\theta}^{(K-1)})\|_2 + \eta \|\Delta (\vec\nabla\mathcal{L})_\mathrm{max}\|_2 \\
        &\leq \eta K \|\Delta (\vec\nabla\mathcal{L})_\mathrm{max}\|_2,
    \end{aligned}
\end{equation}
where we used that the error at the initial point is zero, $\|\Delta(\vec{\delta\theta}^{(0)}\|_2 = 0$.
The error in the loss function gradient can then be written as
\begin{equation}
    \begin{aligned}
        \|\Delta(\vec\nabla\mathcal{L}(\vec{\delta\theta}))\|_2 &= \left\|\frac{\Delta (\vec\nabla F(\vec\theta, \vec\theta + \vec{\delta\theta}))}{2} + \delta\tau \Delta \vec{b}(\vec\theta)  \right\|_2 \\
        &\leq \frac{\|\Delta (\vec\nabla F(\vec\theta, \vec\theta + \vec{\delta\theta}))\|}{2} + \delta\tau \|\Delta \vec{b}(\vec\theta)\|_2 \\
        &\leq \frac{\|\Delta (\vec\nabla F)_\mathrm{max}\|}{2} + \delta\tau \|\Delta\vec b_\mathrm{max}\|_2,
    \end{aligned}
\end{equation}
where $\Delta (\vec\nabla F(\vec\theta, \vec\theta + \vec{\delta\theta})) = \widetilde{\vec\nabla F}(\vec\theta, \vec\theta + \vec{\delta\theta}) - \vec\nabla F(\vec\theta, \vec\theta + \vec{\delta\theta})$ and $\|\Delta (\vec\nabla F)_\mathrm{max}\|_2$ is an upper bound on the maximum fidelity gradient error for any parameter.

The error in the gradient of $F$ can be derived via the parameter-shift rule, as
\begin{equation}
    \begin{aligned}
        |\Delta \partial_i F| &= \frac{\Delta F^{(+)}_i - \Delta F^{(-)}_i}{2}  \\
        &= \mathcal{O}\left(\sqrt{\frac{\mathrm{Var}(F^{(+)}_i)}{N}} + \sqrt{\frac{\mathrm{Var}(F^{(-)}_i)}{N}}\right) \\
        &= \mathcal{O}\left(\frac{1}{\sqrt{N}}\right),
    \end{aligned} 
\end{equation}
where we used that the variance of the fidelity can be bounded for any state $\ket{\psi}$ as 
\begin{equation}
    \begin{aligned}
        \mathrm{Var}(F) = \braket{\psi|P_0^2|\psi} - \braket{\psi|P_0|\psi}^2 
        = \braket{\psi|P_0|\psi} - \braket{\psi|P_0|\psi}^2  
        = F(1 - F) 
        \leq \frac{1}{4}.
    \end{aligned}
\end{equation}
Hence the total error of the fidelity gradient in $\ell_2$ norm is
\begin{equation}
    \|\Delta(\vec\nabla F)_\mathrm{max}\|_2 = \mathcal{O}\left( \sqrt{\frac{d}{N}} \right).
\end{equation}

The bound on $\|\Delta \vec b_\mathrm{max}\|_2$ is already derived in the previous subsection, which gives then a total of
\begin{equation}
    \|\Delta(\vec{\delta\theta})\|_2 = \mathcal{O}\left(\frac{\sqrt{d}K(1 + \delta\tau E_\mathrm{max})}{\sqrt{N}} \right).
\end{equation}
Using the definition $\dot{\vec\theta} = \vec{\delta\theta} / \delta\tau$ we then obtain
\begin{equation}
    \varepsilon_S \leq \Delta_t \sqrt{\lambda_\mathrm{max}} \frac{\|\Delta(\vec{\delta\theta})\|_2}{\delta\tau}
    = \mathcal{O}\left( \sqrt{\frac{\lambda_\mathrm{max} d}{N}}\frac{\Delta_t K (1 + \delta\tau E_\mathrm{max})}{\delta\tau} \right).
\end{equation}

\section{Imaginary-time evolution of the Heisenberg model}\label{app:imag_heisen}

\subsection{Circuit diagram}\label{app:su2_circuit}

\begin{figure}[htp]
    \centering
    \includegraphics[width=0.4\textwidth]{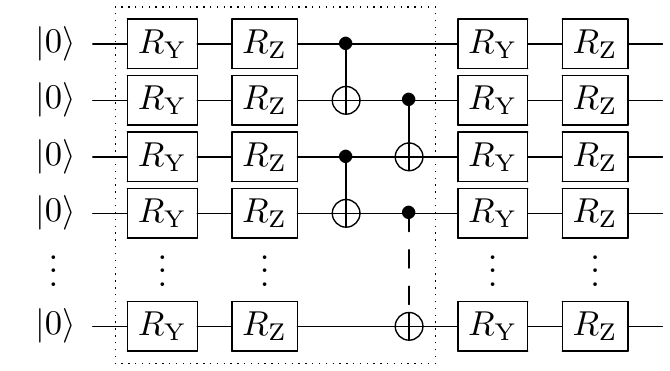}
    \caption{The hardware efficient ansatz for the imaginary-time evolution experiments.
    In the QMETTS experiments the Pauli-$Y$ rotations is replaced by Pauli-$X$ rotations,
    if the evolution starts in the $Y$ basis states $\ket{\pm i}$.
    }
    \label{fig:efficient_su2}
\end{figure}

The circuit used as variational model is schematically presented in Fig.~\ref{fig:efficient_su2}. Each Pauli rotation gate has an independent parameter and the dotted box is repeated several times. For $r$ repetitions and $n$ qubits the total number of tunable parameters is thus $2n(r+1)$. The CNOT entangling gates are arranged in a pairwise manner to minimize the total depth to 2 per entangling layer.

\subsection{Termination and warmstarting}\label{app:warmstarting}

Termination criteria for gradient descent algorithms are typically defined as achieving a minimal threshold in the difference in the loss function between update steps or in the gradient norm.
However, if only noisy readout of the loss function is available these criteria become unreliable as the noise in the evaluation might prevent the termination criterion to be fulfilled even though the algorithm converged. 

One possible resolution would be to consider a moving average over a past batch of iterations. However, depending on the level of noise, this could require a large batchsize and therefore many iterations until the termination can be checked. 
Since the dual time evolution only has to compute small corrections, if small timesteps are performed and the optimization are warmstarted, 
we only expect a few iterations and a moving average is not a resource-efficient solution.
Therefore we use a heuristic where the first optimization uses a large number of steps and the subsequent ones perform a fixed number of few iterations.

\begin{figure}[htp]
    \centering
    \includegraphics[width=0.47\textwidth]{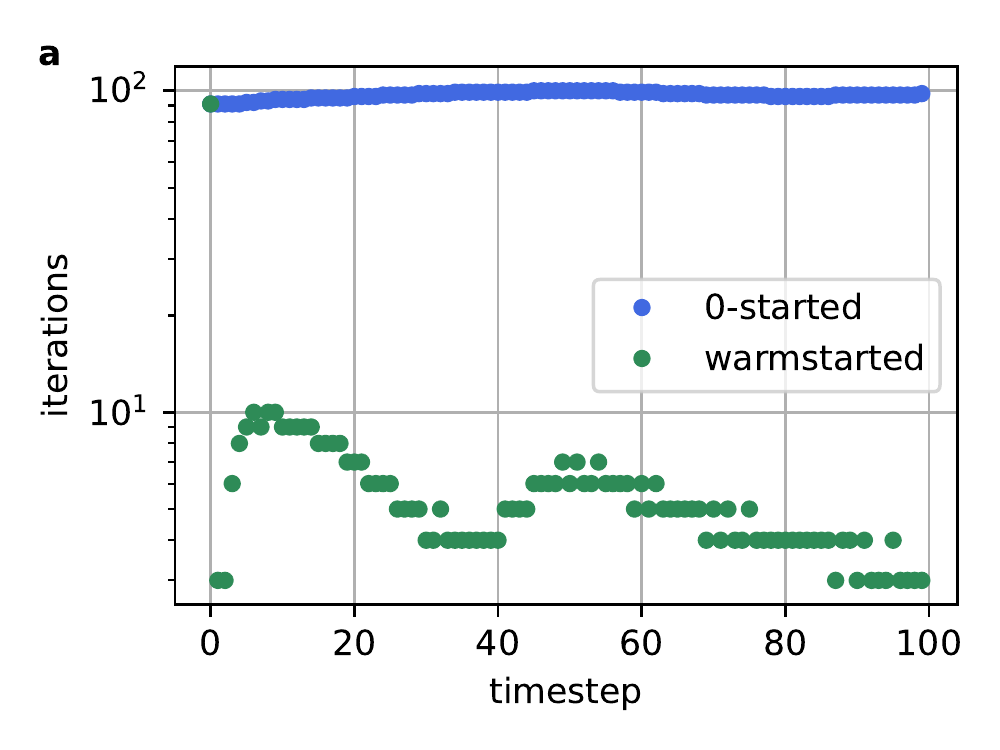}
    \includegraphics[width=0.47\textwidth]{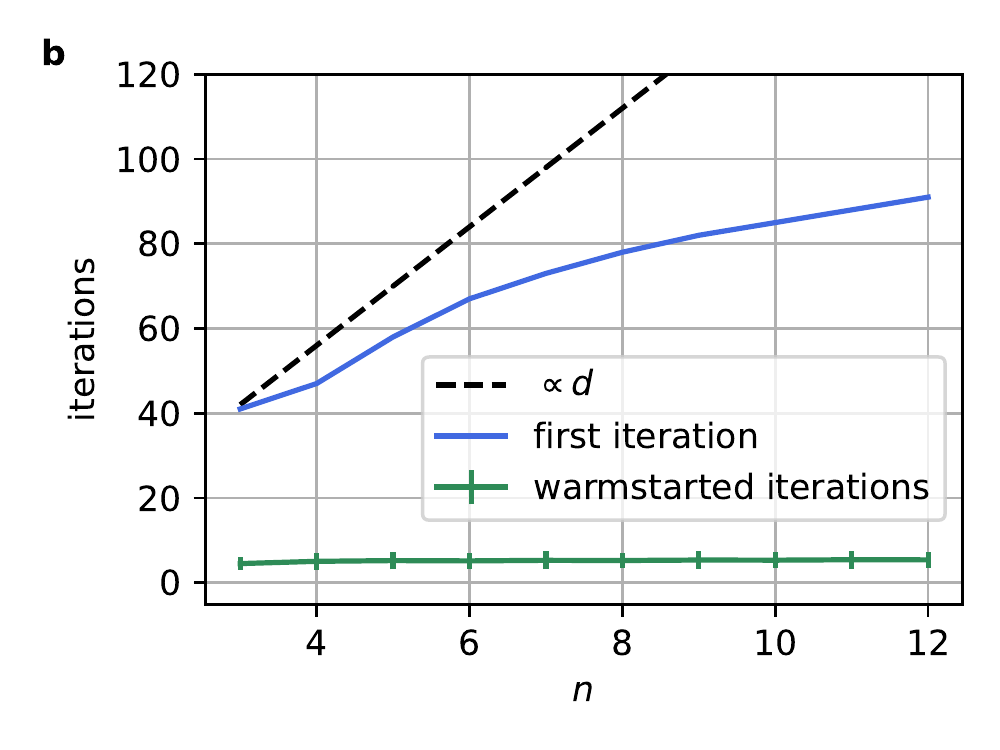}
    \caption{(a) The number of iterations required per timestep until convergence is reached with different initialization techniques.
    (b) The number of iterations for different numbers of qubits and the first iteration and warmstarted iterations. The warmstarted points show mean and standard deviation of the number of iteration of all steps after the first. 
    }
    \label{fig:circuit_requirements}
\end{figure}

To demonstrate the effectiveness of warmstarting and to calibrate the number of required steps for noisy evaluations we investigate the dual time evolution in an ideal setting with exact statevector simulations and no finite-sampling statistics.
First, we perform the time evolution for a Heisenberg Hamiltonian with periodic boundary conditions with $n=12$ sites, $J=1/4$, $g=-1$ and the initial state $\ket{+}^{\otimes  n}$. As circuit model we use the hardware efficient circuit from Fig.~\ref{fig:efficient_su2} with $r=6$ repetitions and optimize the update step with a gradient descent routine with a fixed learning rate of $\eta=0.1$. 
In each timestep we iterate until the change in loss function $\Delta\mathcal{L}$ is below the threshold of $10^{-4}\Delta_t = 10^{-6}$.
The results are presented in Fig.~\ref{fig:circuit_requirements}(a), and we observe that warmstarting drastically reduces the number of required iterations until the convergence criterion is reached.

In a second experiment we analyze how the required number of optimization steps scales with the system size. In Fig.~\ref{fig:circuit_requirements}(b) we repeat the above experiment for $n=3$ to 12 spins and track the number of steps in the first iteration and the mean and standard deviation of the warmstarted iterations.
We see that the number of steps scales sublinearly in the number of parameters $d$ and is almost constant for the warmstarted iterations. 

\subsection{Resource requirements for the dual time evolution}\label{app:dual_resources}

This section shows the detailed VarQITE and DualQITE settings for the resource estimations in Sec.~\ref{sec:imaginary}.
For VarQITE we only varied the number of shots and in the dual method we additionally allowed to vary the number of iterations in the optimization in each time step. Especially DualQITE has a lot of additional degrees of freedom that could be optimized, such as the kind of optimizer, in addition to settings shared with VarQITE, such as timestep size. 

Table~\ref{tab:resources} shows the settings for VarQITE and DualQITE for the resource estimation in Fig.~\ref{fig:heisen}(b)
and Table~\ref{tab:resources_sizescaling} the settings for the scaling with system size in Fig.~\ref{fig:sizescaling}.

\begin{table}[htp]
    \centering
    \begin{subtable}{0.49\textwidth}
    \centering
    \begin{tabular}{c|c|c}
        $I_B$ & shots & $N$ \\ \hline
        1.601 & 100 & $\sim 10^8$ \\
        0.558 & 1024 & $\sim 10^9$ \\
        0.149 & 8192 & $\sim 8 \cdot 10^9$
    \end{tabular}
    \caption{Settings for VarQITE.}
    \end{subtable}
    \hfill
    \begin{subtable}{0.49\textwidth}
    \centering
    \begin{tabular}{c|c|c|c|c}
        $I_B$ & shots & $K_0$ & $K_{>0}$ & $N$ \\ \hline
        0.937 & 100 & 100 & 10 & $\sim 2.5 \cdot 10^7$ \\
        0.735 & 100 & 200 & 20 & $\sim 5 \cdot 10^7$ \\
        0.305 & 1024 & 100 & 10 & $\sim 2.5 \cdot 10^8$ \\
        0.236 & 1024 & 200 & 20 & $\sim 5 \cdot 10^8$ \\
        0.153 & 2048 & 250 & 25 & $\sim 10^9$
    \end{tabular}
    \caption{Settings for DualQITE.}
    \end{subtable}
    \caption{Detailed settings for the resource comparison of VarQITE and DualQITE at fixed number of qubits $n=12$: the achieved Bures distance $D_B$, the number of shots per circuit and the total number of measurements $N$. The dual method additionally shows the number of iterations $K_0$ in the first optimization and $K_{>0}$ in the subsequent, warmstarted optimizations. Each optimization used gradient descent with a learning rate of $\eta = 0.1$.}
    \label{tab:resources}
\end{table}

\begin{table}[htp]
    \centering
    \begin{subtable}{0.49\textwidth}
    \centering
    \begin{tabular}{c|c|c}
        $n$ & shots & $N$\\ \hline
        4 & 500 & $4.2 \cdot 10^7$ \\
        6 & 1500 & $4.2 \cdot 10^8$ \\
        8 & 2500 & $1.2 \cdot 10^9$ \\
        10 & 6000 & $6.7 \cdot 10^9$ \\
        12 & 8000 & $1.3\cdot 10^{10}$ 
    \end{tabular}
    \caption{Settings for VarQITE.}
    \end{subtable}
    \hfill
    \begin{subtable}{0.49\textwidth}
    \centering
    \begin{tabular}{c|c|c|c|c|c}
        n & shots & $K_0$ & $K_{>0}$ & $\eta$ & $N$\\ \hline
        4 & 500 & 100 & 15 & 0.07 & $8.8 \cdot 10^7$\\
        6 & 600 & 200 & 25 & 0.07 & $3.3 \cdot 10^8$\\
        8 & 1000 & 100 & 20 & 0.1 & $6 \cdot 10^8$\\
        10 & 1500 & 200 & 25 & 0.12 & $1.7 \cdot 10^9$\\
        12 & 2500 & 200 & 25 & 0.1 & $3.5 \cdot 10^9$\\
        14 & 3000 & 250 & 25 & 0.12 & $4.9 \cdot 10^9$
    \end{tabular}
    \caption{Settings for DualQITE.}
    \end{subtable}
    \caption{Algorithm settings for the size scaling experiments of VarQITE and DualQITE.}
    \label{tab:resources_sizescaling}
\end{table}

\subsection{Gradient benchmark}\label{app:vanishing_gradients}

In this section, we measure how the norm of the loss function gradient, defined as
\begin{equation*}
    \vec\nabla_{\vec{\delta\theta}} \mathcal{L}(\vec\theta) = -\frac{\vec\nabla_{\vec{\delta\theta}} F(\vec\theta, \vec\theta+\vec{\delta\theta})}{2} - \delta\tau \cdot \vec{b}(\vec\theta)
\end{equation*}
scales with the number of qubits $n$ in the Heisenberg Hamiltonian.
This Hamiltonian is 2-local and since the ansatz depth grows logarithmic with the number of qubits we do not expect exponentially vanishing gradients for the evolution gradient $b$ \cite{cerezo_cost-induced_2021}. 
Further, as discussed in the main text, the initial point of each timestep $\vec{\delta\theta}$ is close to $\vec 0$, which ensures the circuit required to measure to fidelity gradient is close to the identity and we do not expect to encounter a barren plateau~\cite{grant_bp-initialization_2019}.

\begin{figure}
    \centering
    \includegraphics[width=\textwidth]{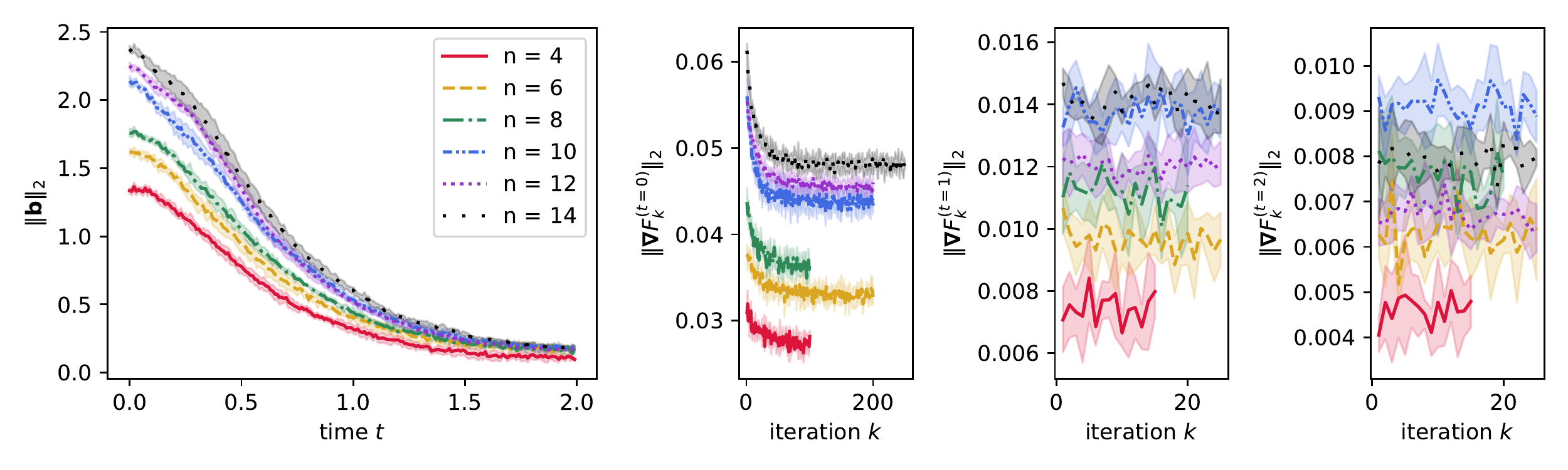}
    \caption{The $\ell_2$ norms for increasing number of qubits $n$ of the imaginary evolution gradient, $\|\vec b\|_2$, during the entire evolution and the norm of the fidelity gradients $\|\vec\nabla F^{(t)}_k\|_2$ at selected times $t$, where $k$ indicates the iteration in the optimization within the timestep. Lengths for the fidelity gradients differ as the optimization were performed using a different number of steps, see Table~\ref{tab:resources_sizescaling}.
}
    \label{fig:gradscaling}
\end{figure}

Since both parts of the loss functions are not in a barren plateau setting, we expect the loss function to be measurable efficiently. In Fig.~\ref{fig:gradscaling}, we measure the $\ell_2$ norm of both the evolution gradient and fidelity gradients and we find that neither gradient vanishes exponentially.
Instead, the evolution gradient increases with system size, which reflects the extensiveness of energy in the Heisenberg model. 
Since we perform imaginary time evolution the system converges towards the ground state and we expect the energy gradients to vanish, once converged. In this case, we do not infer a barren plateau as the gradient norm does not systematically decrease faster for larger systems.
Similarly, the fidelity gradients are expected to decay as the optimal parameter update $\vec{\delta\theta}$ is found.
Note that for system sizes of 12 qubits the exponential decay of the gradients is typically clearly visible~\cite{mcclean_barren_2018}.

\section{Real-time evolution of the Heisenberg model}
\label{app:real_heisen}

\subsection{Circuit diagram}

\begin{figure}[htp]
    \centering
    \includegraphics[width=0.49\textwidth]{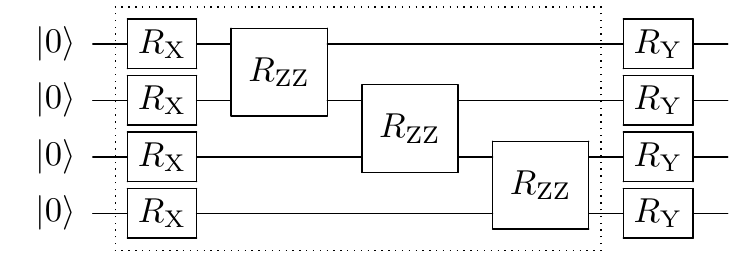}
    \caption{The circuit model used for the real-time evolution experiments.}
    \label{fig:alternating_ansatz}
\end{figure}

In the real-time evolution of the Heisenberg model we use the circuit sketched in Fig.~\ref{fig:alternating_ansatz}, which is the same model used in Ref.~\cite{barison_pvqd_2021}.
The dotted box is repeated three times and the rotation layer alternates between Pauli-$X$ rotations (starting from the first layer) and Pauli-$Y$ rotations.

\end{document}